\documentclass[
reprint,
superscriptaddress,
amsmath,
amssymb,
aps,
prb,
floatfix
]{revtex4-2}

\usepackage{graphicx}
\usepackage{amsmath}
\usepackage{dcolumn}
\usepackage{bm}
\usepackage{xcolor}
\usepackage{hyperref}
\usepackage{comment}
\usepackage{braket}
\usepackage[version=4]{mhchem}
\usepackage{xparse}
\usepackage{physics}
\usepackage{amsfonts}
\usepackage{caption}
\usepackage{subcaption}

\hypersetup{
colorlinks=false,
linktocpage=true,
citebordercolor=[rgb]{0.5,0.5,1.0},
linkbordercolor=[rgb]{1.0,0.5,0.5},
urlbordercolor=[rgb]{0.5,0.5,1.0}
}

\input{macros.sty}

\date{\today}

\begin{document}

\title{Theory of the Spinon-Mediated Witness Spin Glass in Herbertsmithite}

\author{Mitikorn Wood-Thanan}
\affiliation{School of Physics and Astronomy, Cardiff University, Cardiff CF24 3AA, United Kingdom}
\affiliation{School of Physics, University of Bristol, Bristol BS8 1TL, United Kingdom}
\author{Felix Flicker}
\affiliation{School of Physics, University of Bristol, Bristol BS8 1TL, United Kingdom}

\begin{abstract}
Herbertsmithite is a prototypical candidate quantum spin liquid (QSL), believed to feature a long-range entangled ground state and deconfined fractionalised spinon excitations. Confirmation of these properties has been hindered by the presence of magnetic impurities (spin-1/2 Cu$^{2+}$ spins substituted onto non-magnetic Zn$^{2+}$ sites). Recently these impurities were reconceptualised as `witnesses' of the QSL, inheriting long-range interactions and entanglement mediated by the QSL spinons, leading to spin glass formation amongst witnesses below $260\,$mK. Here we present a full theoretical account of this idea. Despite having only one free parameter (the witness-kagome spin coupling), our model captures the full range of experimental data, including: the formation of a spin glass amongst witnesses; the frequency and temperature dependence of the magnetic noise; a sharp peak in the DC magnetic susceptibility as a function of temperature; and the static neutron scattering structure factor at $2\,$K. Both candidate QSLs ($\mathbb{Z}_2$ and $U(1)$) give similar agreement with all data; however, our model predicts a qualitative difference between the $\mathbb{Z}_2$ and $U(1)$ neutron scattering intensities below $260\,$mK, providing a long-sought definitive test to distinguish the two cases. We also predict phase diagrams as a function of temperature and witness concentration, finding a phase transition to different long-range ordered witness states for $\mathbb{Z}_2$ and $U(1)$ at high concentration.
\end{abstract}

\maketitle

%
\section{Introduction}
\label{sec:introduction}
%

Herbertsmithite (ZnCu$_{3}$(OH)$_6$Cl$_2$) is a leading candidate to host a quantum spin liquid (QSL)~\cite{BertEA07,MendelsBert10,PhysRevLett.104.147201,Norman16,BroholmEA20}. This long-sought state of matter is characterised by the lack of magnetic order at absolute zero; long-range many-body entanglement; and deconfined fractionalised `spinon' excitations~\cite{SavaryBalents17,Balents10,ZhouKanodaNg17,
BroholmEA20,KnolleMoessner19,Wen02}. A number of features of Herbertsmithite make it an excellent candidate to host this exotic quantum state of matter. In particular, it features quantum spin-$1/2$ Cu${}^{2+}$ ions on a kagome lattice with an antiferromagnetic (AFM) nearest neighbour exchange energy $J\approx -180\,$K~\cite{ShoresEA05,HeltonEA07,MisguichSindzingre07,MendelsBert10}. These spins are expected to form a Kagome Heisenberg Antiferromagnet (KHAF), which numerical methods such as exact diagonalisation and DMRG predict to host either a $U(1)$ or $\mathbb{Z}_2$ QSL~\cite{RanEA07,HeEA17,YanHuseWhite11,LuRanLee11,DepenbrockEA12} (although more recent machine learning methods have brought this into question~\cite{DuricEA25}). Evidence for the QSL in Herbertsmithite includes inelastic neutron scattering, which reveals a broad continuum of low-energy excitations consistent with spinons~\cite{HeltonEA07,deVriesEA09,HanEA12,Norman16}. 

\begin{figure}
    \includegraphics[width=\columnwidth]{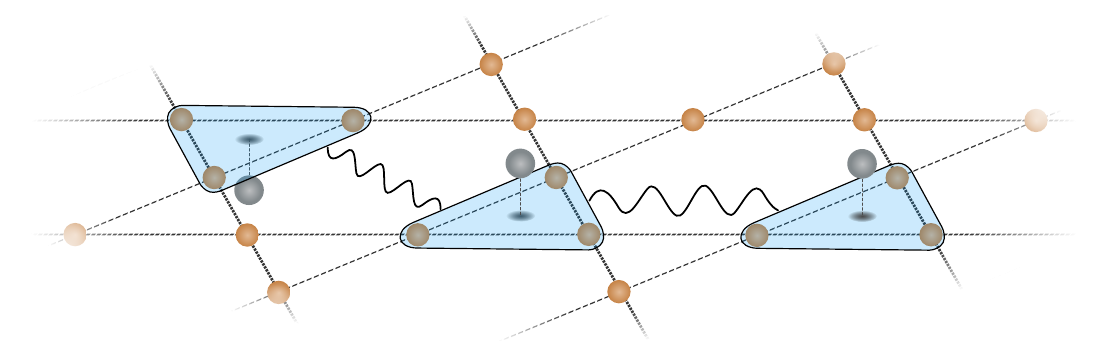}
    \caption{Schematic illustration of our model. Spin-1/2 Cu${}^{2+}$ ions (grey), termed witnesses, randomly occupy a fraction $x$ of non-magnetic Zn sites (not shown). The quantum spin liquid in the kagome Cu${}^{2+}$ plane (brown) mediates an interaction between these witnesses, leading to their observed spin glass formation below $T^*=260\,$mK~\cite{TakahashiEA26}.}
    \label{fig:kagome}
\end{figure}

However, definitive and unambiguous evidence of the QSL in Herbertsmithite remains lacking. A significant complicating factor is the presence of magnetic impurities. Hypothetically, the Cu${}^{2+}$ kagome planes are separated by triangular planes of non-magnetic Zn${}^{2+}$ ions, meaning the kagome layers are isolated 2D systems. In reality, 15-33\% of the Zn${}^{2+}$ ions are substituted with spin-1/2 Cu${}^{2+}$ ions~\cite{LeeEA07,deVriesEA08,FreedmanEA10}. This could either mean that real Herbertsmithite crystals are non-stoichiometric (with an excess of Cu compared to the ideal composition), or that Zn ions substitute onto the Cu planes. These magnetic impurities are known to dominate the observed magnetic behaviour at low temperatures. For example, DC magnetic susceptibility below around $30\,$K is consistent with a signal coming entirely from impurities~\cite{BertEA07}, while inelastic neutron scattering at $2\,$K shows broad features away from high-symmetry points which can be explained by AFM correlations, without long-range order, between nearest neighbour impurities~\cite{OlariuEA08,HanEA16}.

Recent experiments, to which the present authors provided theoretical support, identified previously unseen behaviour in Herbertsmithite at low temperatures~\cite{TakahashiEA26}. Decreasing the temperature below $T^*=260\,$mK, the DC magnetic susceptibility $\chi(T)$ shows a sharp cusp. Below $T^*$ the noise spectrum of magnetization fluctuations as a function of frequency $f$ shows clear $1/f$ behaviour (the noise magnitude rapidly disappears when heating above $T^*$). Along with a lack of any indication of the onset of long-range magnetic order in earlier neutron scattering measurements at $2\,$K, this behaviour pointed towards spin glass formation amongst impurities -- recast as `witnesses' in Ref.~\onlinecite{TakahashiEA26} (nomenclature we adopt here). The spin glass behaviour was confirmed by observation of ageing phenomena: quenching from $400\,$mK to temperatures below $T^*$, there is an excess of magnetic noise (above the $1/f$ signal, and in violation of the fluctuation-dissipation theorem) at the lowest frequencies (corresponding to timescales of hours to days). This additional low-frequency noise power decays with time. 

In this paper we present an expanded account of our theoretical model in Ref.~\onlinecite{TakahashiEA26} that explained these observations. Taking inspiration from the proposal that spinons can mediate an RKKY interaction~\cite{LeggBraunecker19} (in which fermions mediate a long-range interaction between dilute magnetic impurities), we model the interactions between Cu${}^{2+}$ `witnesses' as resulting from the QSL in the kagome layers. The basic picture is presented in Fig.~\ref{fig:kagome}: each witness, a copper ion sitting in a zinc layer, couples to the three copper ions in the kagome layers above and below. Each of these kagome copper ions couples to all other copper ions in the same kagome layer, owing to the long-range entanglement of the QSL. This leads to an indirect coupling between any pair of witness spins touching the same kagome layer. 

We obtain the effective witness-witness interactions $J_{ij}$ (where $i,j$ are witness sites) by integrating out the QSL using the kagome spin-spin susceptibility using a spinon-fermion mean field decomposition. To proceed, we use a simplified model in which the witness quantum spins-1/2 are reduced to classical Ising spins. This step is purely pragmatic, allowing us to use large-scale classical Monte Carlo methods to simulate witness correlations and dynamics. Despite the simplification, we find an excellent agreement with the full range of experimental results. We consider other possible models of witness interactions, including near-neighbour direct exchange and magnons; while some capture many of the results, none is compatible with all the observations. 

We consider both candidate QSLs in Herbertsmithite: the gapped $\mathbb{Z}_2$ QSL, and the $U(1)$ QSL which has a Dirac Fermi surface. We find that both give qualitatively similar results. The $\mathbb{Z}_2$ QSL gives a quantitatively better match to the experimental data in every case, but the approximations in our model (such as classical witness spins and mean field QSL treatment) likely negate any meaningful difference. 

Turning to new predictions, however, we find a qualitative difference between the behaviours of the candidate QSLs. While both QSLs give identical static neutron scattering patterns at $2\,$K, with broad features as observed experimentally, we predict that below $260\,$mK (we look at $100\,$mK) the patterns are entirely different. New broad features develop in the $U(1)$ QSL compared to the $\mathbb{Z}_2$ QSL. By clearly distinguishing between the candidate QSLs, such a measurement could provide the definitive evidence of the QSL itself.

We additionally make new predictions for the evolution of the spin glass as a function of witness concentration $x$. Above a certain threshold of $x$, the spin glass is replaced by a long-range ordered (LRO) phase. We find different long-range orders and correspondingly different threshold $x$ for $U(1)$ and $\mathbb{Z}_2$. While such concentrations cannot be reached physically, the forms of the hypothetical long-range order provide insight into the nature of the spin glass states at the natural concentration. In particular, we argue that the structure observed in neutron scattering~\cite{HanEA16} can be understood as (dynamically glassy) frustrated short-range ordered regions.

This paper proceeds as follows. In Sec.~\ref{sec:Model} we introduce our model of QSL-mediated interactions between witness spins. We outline certain basic assumptions and definitions in Sec.~\ref{subsec:assumptions}. We detail the witness-witness interaction Hamiltonian in Sec.~\ref{subsec:witness_interactions} and the $\mathbb{Z}_2$ and $U(1)$ QSL Hamiltonians in Sec.~\ref{subsec:QSLs}. In Sec.~\ref{subsec:jij} we integrate out the QSLs to derive the effective witness-witness interactions $J_{ij}$. In Sec.~\ref{sec:Methods} we detail our numerical Methods, specifically the large-scale classical Monte Carlo simulations of witness spins. In Sec.~\ref{sec:Results} we present our results, showing a match to neutron scattering (Sec.~\ref{subsec:neutrons}), magnetic susceptibility (Sec.~\ref{subsec:chi_T}), the spin glass order parameter (Sec.~\ref{subsec:q}), and the noise spectral density (Sec.~\ref{subsec:PSD}). We make new predictions for the evolution of the behaviour with varying witness concentration in Sec.~\ref{subsec:x}. In Sec.~\ref{sec:conclusions} we present our conclusions.

%
\section{Model}
\label{sec:Model}
%

The basic assumption of our model is that witness interactions are mediated by the quantum spin liquid in the kagome layers. The picture to hold in mind is that of the RKKY interaction between magnetic impurities in the presence of a Fermi surface: clear evidence of a Fermi surface in Herbertsmithite would point towards the QSL, since the material is a band insulator. This story is complicated, however, by the fact that neither of the two candidate QSLs in Herbertsmithite has a spinon Fermi surface. The $U(1)[0,\pi]$ has pointlike Dirac nodes at the Fermi level~\cite{RanEA07,HermeleEA08}, while the $\mathbb{Z}_2[0,\pi]\beta$ is gapped~\cite{LuRanLee11}. Nevertheless, an RKKY-type calculation can be performed, leading to different forms of interaction in the two cases.

\subsection{Assumptions and definitions}
\label{subsec:assumptions}

We make the following assumptions on the nature of the impurities in Herbertsmithite.
\begin{itemize}
\item A proportion, $x$, of Zn sites are occupied by magnetic Cu${}^{2+}$ ions with no corresponding defects in the kagome copper plane. Following the analysis in Ref.~\onlinecite{TakahashiEA26}, we set $x=0.33$ unless otherwise mentioned.
\item There is no back-action of the witnesses on the kagome copper spins. 
\item Witnesses do not induce any coupling between kagome planes.
\end{itemize}

We label witness sites (which live in Zn layers) $i,j$ and kagome sites (in Cu layers) $l,m$. 

\subsection{Witness-Witness interactions}
\label{subsec:witness_interactions}

The idea of a spinon-mediated RKKY interaction was introduced in Ref.~\onlinecite{LeggBraunecker19} and Ref.~\onlinecite{DhochakEA10}. We treat the witnesses as dilute, and assume that each witness couples to the three kagome spins nearest to it in each of its two neighbouring kagome layers, via an interaction Hamiltonian
\begin{align}
    \hat{H}_{WK}=\gamma \hat{s}_W\cdot\hat{s}_K
\end{align}
where $\hat{s}_W$ is the witness spin and $\hat{s}_K$ is the kagome spin. We assume that this coupling is incoherent: that is, we treat the coupling of the witness to each spin in the plaquette independently and neglect any coupling to the coherent superposition of the three spins together. This assumption is a conventional approach in the study of RKKY in graphene~\cite{Saremi07,BlackSchaffer10,PowerFerreira13}.

Following the standard RKKY prescription, an interaction is induced between any pair of witnesses neighbouring the same kagome plane. Each witness couples to the three nearest kagome spins, and these kagome spins are coupled by the static spin susceptibility $\chi^0_{lm}$ of the QSL, where $l,m$ are kagome sites. The induced witness interaction is then
\begin{equation}
\label{eq:J_ij}
    J_{ij} = (\gamma^2/4)\sum_{l\in\therefore_i}\sum_{m\in\therefore_j}\chi^0_{lm}
\end{equation}
where the sum is over the three kagome spins neighbouring witness spin $i$ (this triangle of spins being denoted $\therefore_i$). Witnesses then interact via
\begin{align}
    \hat{H}_{WW}=-\sum_{ij}J_{ij}\hat{s}_i\cdot\hat{s}_j
\end{align}
where the sum is over all pairs of witnesses, and $J_{ij}$ is non-zero whenever $i$ and $j$ neighbour the same kagome layer.

The static spin susceptibility can be calculated using 
\begin{equation}\label{eq:sssusceptibility}
\chi^{0}_{lm}=-\frac{1}{\pi}\int_{-\infty}^{E_{F}}\mathfrak{Im}\left[G_{lm}\left(E\right)G_{ml}\left(E\right)\right]\text{d}E
\end{equation}
where $E_F$ is the Fermi energy, and 
\begin{align}
\label{eq:G}
G_{lm}\left(E\right)=\langle l|\left(E+i\eta-\hat{H}_{\text{QSL}}\right)^{-1}|m\rangle
\end{align}
is the real-space Green's function connecting kagome sites $l$ and $m$ at energy $E$ ($\eta$ is a small positive regularisation) and $\hat{H}_{\text{QSL}}$ is the Hamiltonian describing the QSL. 

We proceed to detail the forms of the two candidate QSL Hamiltonians in Herbertsmithite.

\subsection{Candidate QSLs in Herbertsmithite}
\label{subsec:QSLs}

We make the standard assumption that the kagome layers in Herbertsmithite are well represented by the KHAF model
\begin{align}
\hat{H}_{KK}=-J\sum_{\langle lm\rangle}\hat{s}^{K}_{l}\cdot\hat{s}^{K}_{m}
\end{align}
where $J\approx -180\,$K~\cite{MendelsBert10}. 

\begin{figure}
    \centering
\includegraphics[width=\linewidth]{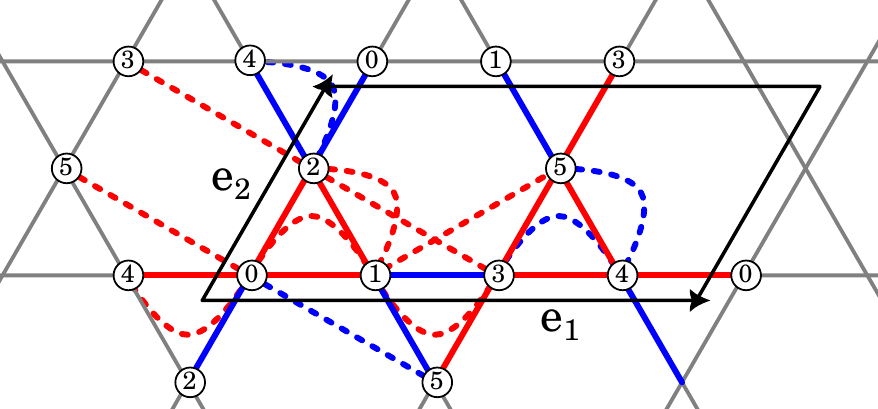}
\caption{The signs of the hopping parameters used to define the QSLs. Solid lines indicate nearest neighbour hoppings $t^{(1)}_{lm}$ (red positive, blue negative) which ensure $\pi$ gauge flux through each hexagon (the product of hoppings around a hexagon is $-1$) and 0 flux through triangles. Dashed lines indicate second-neighbour hoppings $t^{(2)}_{lm}$ used to define the $\mathbb{Z}_2$ QSL. The 6-atom unit cell, and basis vectors $e_1$ and $e_2$, are also shown.}
    \label{fig:QSL_hoppings}
\end{figure}

The Hamiltonian governing QSLs in the KHAF can be obtained by making a Schwinger fermion decomposition~\cite{RanEA07}
\begin{align}
    \hat{\mathbf{s}}^K_l=\frac{1}{2}\sum_{\alpha\beta}\hat{f}^\dagger_{l\alpha}{\mathbf\sigma}^{\alpha\beta}\hat{f}^{\phantom{\dagger}}_{l\beta}
\end{align}
where $\hat{f}^\dagger_{l\alpha}$ is a spinon creation operator on kagome site $l$, $\alpha=\{\uparrow,\downarrow\}$ labels the spin, and $\mathbf{\sigma}$ is a vector of Pauli matrices. This decoupling unphysically doubles the size of the Hilbert space; physical states obey the constraint
\begin{align}
    \sum_{\alpha}\hat{f}^\dagger_{l\alpha}\hat{f}^{\phantom{\dagger}}_{l\alpha}=1
\end{align}
on each site $i$.

\begin{figure}
    \centering
\includegraphics[width=\linewidth]{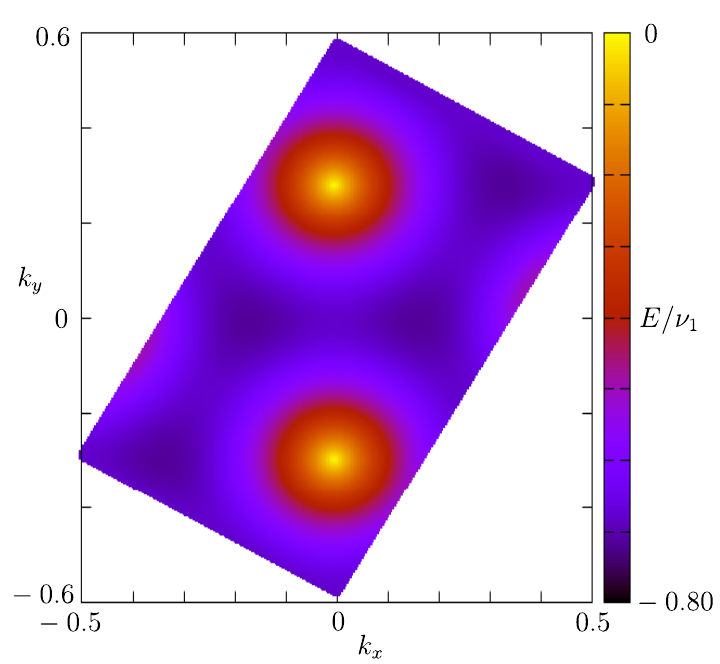}
\caption{The Brillouin zone corresponding to the unit cell convention in Fig.~\ref{fig:QSL_hoppings}; colours indicate the energy of the lowest-energy dispersing spinon band for the $U(1)$ QSL (see Fig.~\ref{fig:bandstructure}), with Dirac nodes evident at $E_F$. Node locations are gauge dependent, but their separation is gauge invariant and sets the behaviour of $J_{ij}$.}
    \label{fig:BZ}
\end{figure}

Under this substitution $\hat{H}_{KK}$ becomes quartic in spinon operators. Using a Hubbard Stratanovitch transformation this can be rewritten as a quadratic in spinon operators at the cost of introducing a new field $\hat{\nu}_{lm}$ living on edges. Taking the mean field values of $\hat{\nu}_{lm}$ then reduces the KHAF to a quadratic Hamiltonian in spinon operators:
\begin{align}
    \hat{H}_{KK}=-J\sum_{\langle lm\rangle}\sum_{\alpha}\nu_{1}t^{(1)}_{lm}\hat{f}^\dagger_{l\alpha}\hat{f}^{\phantom{\dagger}}_{m\alpha}+\textrm{H.c.}
\end{align}
where the mean field value of $\hat{\nu}_{lm}$ is $\nu_{1}t^{(1)}_{lm}$ with unit magnitude hoppings $|t^{(1)}_{lm}|=1$.

Different QSLs can then be classified by the gauge flux passing through each triangle and hexagon in the kagome lattice (equivalently, by the Berry phase picked up by a spinon traversing one of these plaquettes). We proceed to list the two candidate mean field QSLs in Herbertsmithite.

\subsubsection{$U(1)[0,\pi]$ QSL}

The $U(1)[0,\pi]$ QSL has zero flux through triangles and $\pi$ flux through hexagons~\cite{Hastings00,RanEA07,HermeleEA08}. It was previously found to have the lowest energy of all mean field $U(1)$ QSLs in the KHAF~\cite{RanEA07}. The required gauge fluxes can be achieved with real nearest neighbour spinon tight-binding hoppings $t^{(1)}_{lm}$ using a six-site unit cell (a doubling of the minimal structural unit cell, to accommodate the $\pi$ flux) as detailed in Ref.~\onlinecite{RanEA07}. We use the same hoppings and unit cell definitions here, shown in Fig.~\ref{fig:QSL_hoppings}. In these definitions all $t^{(1)}_{lm}$ have the same magnitude, and vary only in sign. The magnitude $\nu_1$, which sets the overall spinon bandwidth, was reported in Ref.~\onlinecite{Hastings00} to be $\nu_1=0.221$.

The resulting spinon bandstructure contains Dirac nodes at the Fermi level $E_F$~\cite{HeEA17} (Fig.~\ref{fig:BZ} and Fig.~\ref{fig:bandstructure}). Different gauge choices for the QSL (different hoppings leading to the same gauge fluxes) cause the Dirac nodes to move relative to the Brillouin zone, but the separation between the Dirac nodes is gauge invariant. This separation sets the RKKY wavevector. 

\subsubsection{$\mathbb{Z}_2[0,\pi]\beta$ QSL}

The set of $\mathbb{Z}_2$ topologically ordered QSLs can be obtained from `parent' $U(1)$ QSLs by introducing further-neighbour couplings so as to induce a mass gap~\cite{LuRanLee11}. Among the offspring of the $U(1)[0,\pi]$ QSL, only one, labelled $\mathbb{Z}_2[0,\pi]\beta$, has a gap. For this reason it has also been proposed as a candidate QSL in Herbertsmithite.

Ref.~\onlinecite{LuRanLee11} sets up the following mean field spinon Hamiltonian:
\begin{widetext}
\begin{align}
\hat{H}_{\mathbb{Z}_2[0,\pi]\beta} & =\sum_{l}^{N}\lambda_{3}\left(\hat{f}_{l\uparrow}^{\dagger}\hat{f}_{l\uparrow}+\hat{f}_{l\downarrow}^{\dagger}\hat{f}_{l\downarrow}\right)+\lambda_{1}\left(\hat{f}_{l\uparrow}^{\dagger}\hat{f}_{l\downarrow}^{\dagger}+\hat{f}_{l\downarrow}\hat{f}_{l\uparrow}\right)\\
 & +\left\{ \sum_{lm}\left(\nu_{1}t_{lm}^{\left(1\right)}+\nu_{2}t_{lm}^{\left(2\right)}\right)\left(\hat{f}_{l\uparrow}^{\dagger}\hat{f}_{m\uparrow}+\hat{f}_{l\downarrow}^{\dagger}\hat{f}_{m\downarrow}-\hat{f}_{l\uparrow}\hat{f}_{m\uparrow}^{\dagger}-\hat{f}_{l\downarrow}\hat{f}_{m\downarrow}^{\dagger}\right)\right.\nonumber\\
 & +\left.\Delta_{2}t_{lm}^{\left(2\right)}\left(\hat{f}_{l\uparrow}^{\dagger}\hat{f}_{m\downarrow}^{\dagger}-\hat{f}_{l\downarrow}^{\dagger}\hat{f}_{m\uparrow}^{\dagger}-\hat{f}_{l\uparrow}\hat{f}_{m\downarrow}+\hat{f}_{l\downarrow}\hat{f}_{m\uparrow}\right)\right\}\nonumber
\end{align}
\end{widetext}
where $t_{lm}^{\left(1\right)}$ are the same nearest neighbour hopping signs as for the $U(1)$ QSL, and $t_{lm}^{\left(2\right)}$ are non-zero only for second-nearest neighbours. There are $N$ sites in the system. The magnitudes of $t_{lm}$ are again fixed to unity. We use the hoppings and unit cell definitions from Ref.~\onlinecite{LuRanLee11}, reproduced in Fig.~\ref{fig:QSL_hoppings}.

We find it convenient to rewrite the diagonal terms using
\begin{align}
\left\{ \hat{f}_{l},\hat{f}_{m}^{\dagger}\right\}  & =\delta_{lm}\\
\left\{ \hat{f}_{l},\hat{f}_{m}\right\}  & =0
\end{align}
to give
\begin{widetext}
\begin{align}
\hat{H} & =N\lambda_{3}+\sum_{l}\frac{\lambda_{3}}{2}\left(\hat{f}_{l\uparrow}^{\dagger}\hat{f}_{l\uparrow}-\hat{f}_{l\uparrow}\hat{f}_{l\uparrow}^{\dagger}+\hat{f}_{l\downarrow}^{\dagger}\hat{f}_{l\downarrow}-\hat{f}_{l\downarrow}\hat{f}_{l\downarrow}^{\dagger}\right)+\frac{\lambda_{1}}{2}\left(\hat{f}_{l\uparrow}^{\dagger}\hat{f}_{l\downarrow}^{\dagger}+\hat{f}_{l\downarrow}\hat{f}_{l\uparrow}-\hat{f}_{l\downarrow}^{\dagger}\hat{f}_{l\uparrow}^{\dagger}-\hat{f}_{l\uparrow}\hat{f}_{l\downarrow}\right)\\
 & +\left\{ \sum_{lm}\left(\nu_{1}t_{lm}^{\left(1\right)}+\nu_{2}t_{lm}^{\left(2\right)}\right)\left(\hat{f}_{l\uparrow}^{\dagger}\hat{f}_{m\uparrow}+\hat{f}_{l\downarrow}^{\dagger}\hat{f}_{m\downarrow}-\hat{f}_{l\uparrow}\hat{f}_{m\uparrow}^{\dagger}-\hat{f}_{l\downarrow}\hat{f}_{m\downarrow}^{\dagger}\right)\right.\nonumber\\
 & +\left.\Delta_{2}t_{lm}^{\left(2\right)}\left(\hat{f}_{l\uparrow}^{\dagger}\hat{f}_{m\downarrow}^{\dagger}-\hat{f}_{l\downarrow}^{\dagger}\hat{f}_{m\uparrow}^{\dagger}-\hat{f}_{l\uparrow}\hat{f}_{m\downarrow}+\hat{f}_{l\downarrow}\hat{f}_{m\uparrow}\right)\right\}\nonumber.
\end{align}
The initial $N\lambda_{3}$ acts as an overall chemical potential and can be dropped. The following basis is then block-diagonal:
\begin{equation}
\!\!\!\!\!\!\!\!\hat{H}=\sum_{lm}\left(\left(\hat{f}_{l\uparrow}^{\dagger},\hat{f}_{l\downarrow}\right),\left(\hat{f}_{l\uparrow},\hat{f}_{l\downarrow}^{\dagger}\right)\right)\left(\begin{array}{cc}
\left(h_{lm}\right) & \left(0\right)\\
\left(0\right) & \left(-h_{lm}\right)
\end{array}\right)\left(\begin{array}{c}
\left(\begin{array}{c}
\hat{f}_{m\uparrow}\\
\hat{f}_{m\downarrow}^{\dagger}
\end{array}\right)\\
\left(\begin{array}{c}
\hat{f}_{m\uparrow}^{\dagger}\\
\hat{f}_{m\downarrow}
\end{array}\right)
\end{array}\right)
\end{equation}
\end{widetext}
where 
\begin{equation}
h_{lm}=\left(\begin{array}{cc}
\frac{\lambda_{3}}{2}\delta_{lm}+\nu_{a}t_{lm}^{a} & \frac{\lambda_{1}}{2}\delta_{lm}+\Delta_{2}t_{lm}^{\left(2\right)}\\
\frac{\lambda_{1}}{2}\delta_{lm}+\Delta_{2}t_{lm}^{\left(2\right)} & -\frac{\lambda_{3}}{2}\delta_{lm}-\nu_{a}t_{lm}^{a}
\end{array}\right)
\end{equation}
(a sum over $a=1,2$ is implicit).

Hence, all the information is contained in the upper matrix:
\begin{equation}
\hat{H}=\sum_{lm}\left(\hat{f}_{l\uparrow}^{\dagger},\hat{f}_{l\downarrow}\right)_{\alpha}h_{lm}^{\alpha\beta}\left(\begin{array}{c}
\hat{f}_{m\uparrow}\\
\hat{f}_{m\downarrow}^{\dagger}
\end{array}\right)_{\beta}.
\end{equation}

While Ref.~\onlinecite{HermeleEA08} sets up the mean field ansatz, we are not aware of work obtaining the self-consistent values. We proceed to establish them here.

The self-consistency conditions are~\cite{HermeleEA08}:
\begin{align}
\Delta_{lm} & =-2\left\langle \hat{f}_{l\uparrow}\hat{f}_{m\downarrow}\right\rangle =2\left\langle \hat{f}_{l\downarrow}\hat{f}_{m\uparrow}\right\rangle \label{eq:sc: Delta}\\
t_{lm} & =2\left\langle \hat{f}_{l\uparrow}^{\dagger}\hat{f}_{m\uparrow}\right\rangle =2\left\langle \hat{f}_{l\downarrow}^{\dagger}\hat{f}_{m\downarrow}\right\rangle \label{eq:sc: chi}\\
0 & =\left\langle \hat{f}_{l\uparrow}\hat{f}_{m\uparrow}\right\rangle =\left\langle \hat{f}_{l\downarrow}\hat{f}_{m\downarrow}\right\rangle =\left\langle \hat{f}_{l\uparrow}^{\dagger}\hat{f}_{m\downarrow}\right\rangle =\left\langle \hat{f}_{l\downarrow}^{\dagger}\hat{f}_{m\uparrow}\right\rangle .
\end{align}
The physical Hilbert space, global half-filling, is enforced by the Lagrange multipliers $\lambda_{1}$ and $\lambda_{3}$:
\begin{align}
\lambda_{1}: & \quad0=\sum_{l}\left\langle \hat{f}_{l\uparrow}\hat{f}_{l\downarrow}\right\rangle -\left\langle \hat{f}_{l\downarrow}\hat{f}_{l\uparrow}\right\rangle \label{eq:sc: fermions}
\end{align}
\begin{equation}
\lambda_{3}:\quad1=\sum_{l}\left\langle \hat{f}_{l\uparrow}^{\dagger}\hat{f}_{l\uparrow}\right\rangle +\left\langle \hat{f}_{l\downarrow}^{\dagger}\hat{f}_{l\downarrow}\right\rangle .\label{eq:1/2-filling}
\end{equation}

Following the standard self-consistent mean field method, we proceed to diagonalise the Hamiltonian
\begin{equation}
\hat{H}=\sum_{lm}\left(\hat{\gamma}_{l1}^{\dagger},\hat{\gamma}_{l2}^{\dagger}\right)D_{lm}\left(\begin{array}{c}
\hat{\gamma}_{m1}\\
\hat{\gamma}_{m2}
\end{array}\right)
\end{equation}
with diagonal $D$, and
\begin{align}
h= & UDU^{\dagger}\\
\left(\begin{array}{c}
\hat{f}_{l\uparrow}\\
\hat{f}_{l\downarrow}^{\dagger}
\end{array}\right)=U_{lm}\left(\begin{array}{c}
\hat{\gamma}_{m1}\\
\hat{\gamma}_{m2}
\end{array}\right)= & \left(\begin{array}{c}
U_{lm}^{11}\hat{\gamma}_{m1}+U_{lm}^{12}\hat{\gamma}_{m2}\\
U_{lm}^{21}\hat{\gamma}_{m1}+U_{lm}^{22}\hat{\gamma}_{m2}
\end{array}\right)
\end{align}
and the Hermitian conjugate gives the other required terms:
\begin{equation}
\left(\hat{f}_{l\uparrow}^{\dagger},\hat{f}_{l\downarrow}\right)=\left(U_{lm}^{11*}\hat{\gamma}_{m1}^{\dagger}+U_{lm}^{12*}\hat{\gamma}_{m2}^{\dagger},U_{lm}^{21*}\hat{\gamma}_{m1}^{\dagger}+U_{lm}^{22*}\hat{\gamma}_{m2}^{\dagger}\right).
\end{equation}

In this basis, 
\begin{align}
\left\{ \hat{\gamma}_{l\alpha},\hat{\gamma}_{m\beta}\right\}  & =\left\{ \hat{\gamma}_{l\alpha}^{\dagger},\hat{\gamma}_{m\beta}^{\dagger}\right\} =0\\
\left\{ \hat{\gamma}_{l\alpha},\hat{\gamma}_{m\beta}^{\dagger}\right\}  & =\delta_{lm}\delta_{\alpha\beta}\\
\left\langle \hat{\gamma}_{l\alpha}^{\dagger}\hat{\gamma}_{m\beta}\right\rangle  & =\delta_{lm}\delta_{\alpha\beta}n_{D}\left(D_{ll}^{\alpha\alpha}\right)
\end{align}
where $n_{D}$ is the Fermi-Dirac distribution. However, note that the eigenvalues $D_{ll}$ are ordered low to high, and the spectrum is symmetric about zero. Hence, at $T\approx0$ (since we are concerned with temperatures below $1\,$K, with  $J=-180\,$K), $n_{D}\left(D_{22}^{\alpha\alpha}\right)=0$ and $n_{D}\left(D_{11}^{\alpha\alpha}\right)=1$.

Entering these expressions into the self-consistency conditions gives
\begin{align*}
\Delta_{lm} & =-2\sum_{l'm'}\left\langle \left(U_{lm'}^{11}\hat{\gamma}_{m'1}+U_{lm'}^{12}\hat{\gamma}_{m'2}\right)\left(U_{ml'}^{21*}\hat{\gamma}_{l'1}^{\dagger}+U_{ml'}^{22*}\hat{\gamma}_{l'2}^{\dagger}\right)\right\rangle \\
 & =\sum_{l'm'}-2U_{lm'}^{11}U_{ml'}^{21*}\left\langle \hat{\gamma}_{m'1}\hat{\gamma}_{l'1}^{\dagger}\right\rangle -2U_{lm'}^{12}U_{ml'}^{22*}\left\langle \hat{\gamma}_{m'2}\hat{\gamma}_{l'2}^{\dagger}\right\rangle \\
 & =\sum_{l'm'}2U_{lm'}^{11}U_{ml'}^{21*}\left\langle \hat{\gamma}_{l'1}^{\dagger}\hat{\gamma}_{m'1}\right\rangle +2U_{lm'}^{12}U_{ml'}^{22*}\left\langle \hat{\gamma}_{l'2}^{\dagger}\hat{\gamma}_{m'2}\right\rangle \\
 & =\sum_{m'}2U_{lm'}^{11}U_{mm'}^{21*}n_{D}\left(D_{m'm'}^{11}\right)+2U_{lm'}^{12}U_{mm'}^{22*}n_{D}\left(D_{m'm'}^{22}\right)\\
 & =2\sum_{m}U_{lm'}^{11}U_{mm'}^{21*}.
\end{align*}

Overall we find:
\begin{align}
\Delta_{2}t_{lm}^{\left(2\right)} & =2\sum_{m'}U_{lm'}^{11}U_{mm'}^{21*}\\
\nu_{a}t_{lm}^{\left(a\right)} & =2\sum_{m'}U_{lm'}^{11*}U_{mm'}^{11}\\
\lambda_{1}:0 & =\sum_{lm}U_{lm}^{12}U_{lm}^{22*}-U_{lm}^{21*}U_{lm}^{11}\\
\lambda_{3}:1 & =\sum_{l}\left|U_{ll}^{11}\right|^{2}+\left|U_{ll}^{22}\right|^{2}.
\end{align}

To proceed, we set $\nu_{1}=1$, defining the energy scale. To find $\Delta_{2}$
and $\nu_{2}$ we use
\begin{align}
h_{01}^{12} & =\Delta_{2}\exp\left(-2\pi i\boldsymbol{k}\cdot e_{2}\right)\\
h_{01}^{11} & =\nu_{1}+\nu_{2}\exp\left(-2\pi i\boldsymbol{k}\cdot e_{2}\right)\\
h_{13}^{11} & =-\nu_{1}+\nu_{2}\exp\left(-2\pi i\boldsymbol{k}\cdot e_{2}\right)
\end{align}
to give
\begin{align}
\Delta_{2} & =h_{01}^{12}\exp\left(2\pi i\boldsymbol{k}\cdot e_{2}\right)\\
\nu_{2} & =\frac{h_{01}^{11}+h_{13}^{11}}{2}\exp\left(2\pi i\boldsymbol{k}\cdot e_{2}\right)
\end{align}
and, under self-consistency,
\begin{align}
\Delta_{2} & =2\exp\left(2\pi i\boldsymbol{k}\cdot e_{2}\right)\sum_{m}U_{1m}^{11}U_{5m}^{21*}\\
\nu_{2} & =\exp\left(2\pi i\boldsymbol{k}\cdot e_{2}\right)\sum_{m}U_{0m}^{11*}U_{1m}^{11}+U_{1m}^{11*}U_{3m}^{11}.
\end{align}

Working in $k-$space at $k=0$ (the gap should be constant in $k$), we find a self-consistent solution with
\begin{align}
\Delta_{2} & =0.4583\\
\nu_{2} & =-0.2849\\
\lambda_{1} & =0.4327\\
\lambda_{3} & =1.500.
\end{align}

\begin{figure}
    \centering
\includegraphics[width=0.9\linewidth]{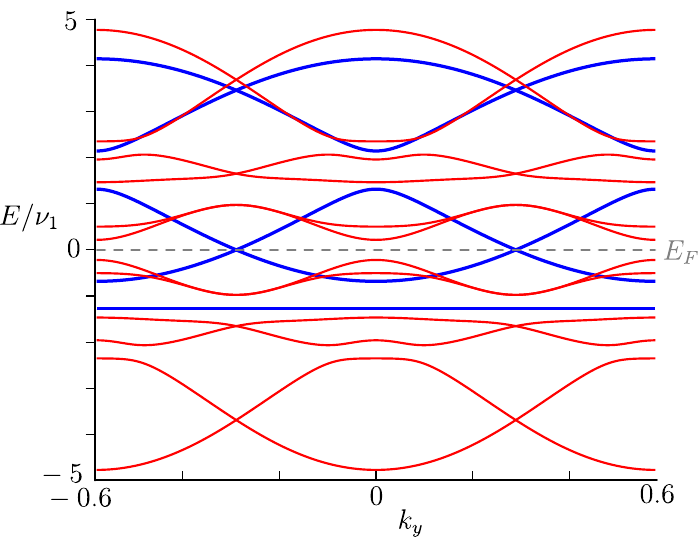}
\caption{Spinon bandstructures along $k_y$ for the $U(1)$ QSL (blue) and $\mathbb{Z}_2$ QSL (red).}
    \label{fig:bandstructure}
\end{figure}

We used a tolerance of $10^{-3}$ in finding the constraints with the Lagrange multipliers:
\begin{align*}
\lambda_{1}\implies & \sum_{lm'}U_{lm'}^{11*}U_{lm'}^{21}=-8.4\times10^{-4}\,\,(\equiv0)\\
\lambda_{3}\implies & \sum_{l}\left|U_{ll}^{11}\right|^{2}+\left|U_{ll}^{22}\right|^{2}=1.001\,\,(\equiv1).
\end{align*}
The resulting spinon bandstructure is shown in Fig.\ref{fig:bandstructure}.

By inspecting the density of states we find a gap of $2\Delta=0.44t_{1}$. This is in good agreement with more exact methods. These are typically given in terms of $J$ rather than $t_1$; the conversion is given as $t_{1}=0.4J$ by Ref.~\onlinecite{RanEA07} and $t_{1}=0.66J$ by Refs.~\onlinecite{Hastings00,GregorMotrunich08}. 

Previous estimates of the $\mathbb{Z}_2[0,\pi]\beta$ gap are $2\Delta/J=0.05\pm0.02$~\cite{FuEA15} (using nuclear magnetic resonance, which may be seeing either the singlet gap or triplet gap); $2\Delta/J=0.095\pm0.025$~\cite{YanHuseWhite11} (Density Matrix Renormalisation Group (DMRG)); $2\Delta/J=0.17$~\cite{LauchliEA19} (exact diagonalisation); and $2\Delta/J=0.13$~\cite{DepenbrockEA12} (DMRG). Our result compares remarkably well with the exact diagonalisation result of Ref.~\onlinecite{LauchliEA19}, when using the conversion factor of Ref.~\onlinecite{RanEA07}, which gives $2\Delta=0.43t_1$.

\subsection{Witness-Witness Susceptibility $J_{ij}$}
\label{subsec:jij}

The witness-witness interaction $J_{ij}$ is set by the kagome-kagome susceptibility as obtained in Sec.~\ref{subsec:QSLs}. The expected form of interactions therefore depends sensitively on the type of QSL.

The standard RKKY interaction occurs between dilute magnetic impurities in the presence of a bulk Fermi surface. It takes the characteristic form\cite{Kittel}
\begin{align}\label{eq:RKKY}
J_{ij}^{\textrm{RKKY}}\propto\frac{\sin(2k_F r_{ij})-2k_F r_{ij}\cos\left(2k_F r_{ij}\right)}{r_{ij}^{D+1}}
\end{align}
where $r_{ij}$ is the separation between impurities at positions $i$ and $j$, and $D$ is the spatial dimension. 

Importantly, however, neither of the candidate QSLs in Herbertsmithite has a bulk spinon Fermi surface. Therefore two other known results for RKKY-type interactions are relevant. The Fermi surface of Graphene takes the form of two Dirac nodes. Since graphene has $D=2$ the general expression in Eq.\ref{eq:RKKY} would suggest a leading-order oscillatory $1/r^2$ behaviour~\cite{Kittel}. In fact, the interaction takes the form
\begin{align}
    \label{eq:RKKY_graphene}
J_{ij}^{\textrm{graphene}}\propto\frac{1}{r_{ij}^3}.
\end{align}
The fast decay comes from the Dirac nodal nature of the Fermi surface, while the lack of oscillations comes from the bipartite nature of the honeycomb lattice~\cite{Saremi07}. Importantly, the interaction is purely antiferromagnetic at all distances. 

Bipartiteness means that the honeycomb lattice sites can be labelled A and B such that As only connect to Bs and vice versa. In Eq.~\eqref{eq:RKKY_graphene} the impurities are assumed to lie at the centres of honeycomb hexagons. There are 6 honeycomb neighbours to each impurity, and so there are 36 honeycomb-honeycomb terms to sum over to get $J_{ij}$. Of these, half are AA or BB couplings and half are AB couplings. Combined with the relative strengths of AA/BB and AB, this results in the non-oscillatory pure AFM interactions of Eq.~\eqref{eq:RKKY_graphene}. We will find a similar, but less perfect, effect in the tripartite kagome lattice (Fig.~\ref{fig:witness-witness_sum}).

The next relevant comparison is to gapped bilayer graphene. In this case the Dirac Fermi surface of graphene, with its associated Fermi velocity $v_F$, has disappeared with the introduction of a small energy gap of size $\Delta$. The spin-spin susceptibility (no longer really an RKKY interaction) leads to a form\cite{Parhizgar13}
\begin{align}
    \label{eq:RKKY_gapped}
    J_{ij}^{\textrm{gapped graphene}}\propto r_{ij}^{-3/2}\exp(-\frac{2\Delta r_{ij}}{\hbar v_F}).
\end{align}
The interaction is again purely antiferromagnetic at all distances. While this is again a consequence of the bipartite lattice structure, we find a similar effect resulting from the tripartite structure of Herbertsmithite.

We proceed to calculate the static spin susceptibility for different QSLs using Eq.~\eqref{eq:sssusceptibility}. We carry out the energy integral numerically. We obtain the real-space Green's functions as a function of energy using exact diagonalisation of finite size clusters in real space; as a check, we also perform an analytic Fourier transform and redo the calculation in $k$-space. The results are identical up to real-space boundary effects.

\subsubsection{$U(1)$ QSL}
\label{sec:jij_u1}

Working in real space we carry out exact diagonalisation of a $32\times 32$ unit cell cluster (with 6 kagome sites per cell). We used $2\times 10^5$ steps in the numerical energy integral, with the regularisation $\eta=10^{-3}\,$meV in Eq.\eqref{eq:G}. The resulting $J_{ij}$ is shown in Fig.~\ref{fig:J_ij_U1}. The interactions are purely antiferromagnetic everywhere; we find the interaction decays with distance as $J_{ij}\propto r^{-3}$. 

At first sight the excellent agreement is surprising: while both 2D, the graphene and Herbertsmithite lattices are different. In particular, Herbertsmithite is not bipartite, although it is \emph{tripartite}, meaning it has three sublattices A, B, C such that A only connects to B and C, and so on.

On closer inspection, though, the arguments presented for graphene in Ref.~\onlinecite{Saremi07} carry over well to Herbertsmithite. The key observation is that, while the standard RKKY interaction is a function of $k_F r_{ij}$, with a pointlike Dirac Fermi surface there is no notion of Fermi wavevector $k_F$. The relevant inverse lengthscale is instead set by $\mathbf{k}_D$, the wavevector separating inequivalent Dirac nodes. The Dirac nodes must lie at high symmetry points in the Brillouin zone. As a result, $\cos(2\mathbf{k}_D\cdot\mathbf{r}_{ij})$ takes special values (integer/half-integer) on all lattice sites. It is this that leads to the complete disappearance of any oscillatory terms in Eq.~\eqref{eq:RKKY_graphene}.

\begin{figure}
    \centering
\includegraphics[width=\linewidth]{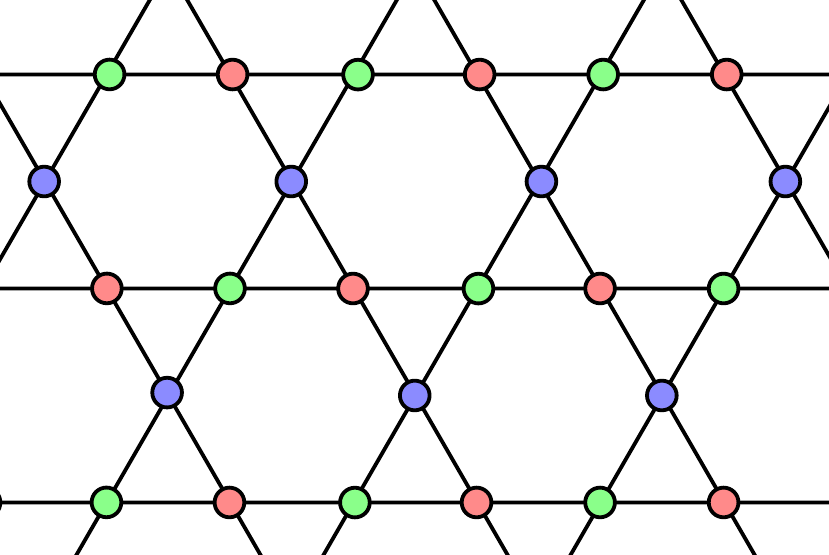}
    \caption{Tripartite colouring of kagome lattice sites. Witnesses sit above/below triangles and couple to the three closest kagome sites. The induced witness-witness coupling $J_{ij}$ involves a sum over 9 kagome spin-spin susceptibilities: 3 of AA/BB/CC type, 6 of AB/BC/AC type.}
    \label{fig:witness-witness_sum}
\end{figure}

Essentially the same arguments work for the $U(1)$ QSL in Herbertsmithite. Here the locations of Dirac nodes in the Brillouin zone are gauge dependent, but the separation between inequivalent nodes, $2\mathbf{k}_D$, is a gauge invariant quantity~\cite{RanEA07}. Just as in graphene, $\cos(2\mathbf{k}_D\cdot\mathbf{r}_{ij})$ takes on special integer/half-integer values on all lattice sites. With witnesses sat a above/below the centres of triangles, there are 9 kagome-kagome susceptibilities to sum over in Eq.~\eqref{eq:J_ij} (see Fig.~\ref{fig:witness-witness_sum}). There are one each of AA, BB, CC, couplings, which we find to be ferromagnetic and weak; then there are a total of six AB, BC, AC couplings which we find to be of equal strength to one another, stronger than the AA type couplings, and antiferromagnetic. Overall, the effect, as in graphene, is to produce a non-oscillatory purely AFM witness-witness interaction. Combined with the random witness site occupation, this frustrating long-range AFM interaction leads to spin glass behaviour.

\begin{figure}
    \centering
\includegraphics[width=\linewidth]{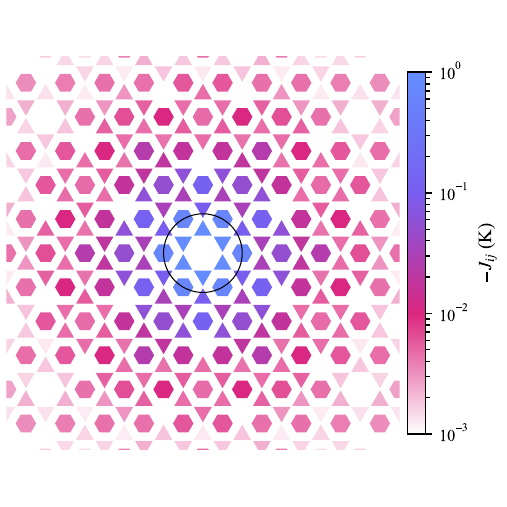}
    \caption{Calculated $J_{ij}$ for the $U(1)$ QSL. Hexagons indicate coupling to witness on the same witness plane and up(down)-triangles indicate coupling to witness on the witness plane above(below) in real space. The circular ring at the center indicates one in-plane atomic spacing.}
    \label{fig:J_ij_U1}
\end{figure}

\subsubsection{$\mathbb{Z}_2$ QSL}
\label{sec:jij_z2}

Working in real space we carry out exact diagonalisation of a $16\times 16$ unit cell cluster. The reduced number compared to the $U(1)$ case is a result of the matrix structure of the $\mathbb{Z}_2$ Hamiltonian.

\begin{figure}
    \centering
    \includegraphics[width=\linewidth]{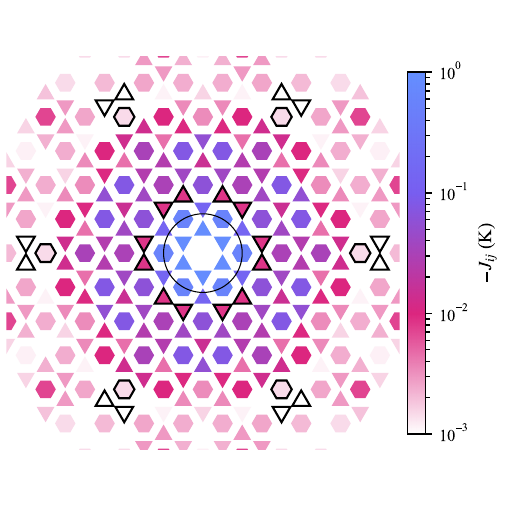}
    \caption{Calculated $J_{ij}$ for the $\mathbb{Z}_2$ QSL. Hexagons indicate coupling to witness on the same witness plane and up(down)-triangles indicate coupling to witness on the witness plane above(below) in real space. The circular ring at the center has a radius of one in-plane atomic spacing. Black borders indicate couplings with opposite sign, \emph{i.e.} ferromagnetic.}
    \label{fig:Z2_hoppings}
\end{figure}

The resulting $J_{ij}$ is shown in Fig.~\ref{fig:Z2_hoppings}. The interactions are almost purely antiferromagnetic as before. The results compare well to the analytic form for gapped graphene, Eq.~\eqref{eq:RKKY_gapped}, with $\Delta/v_F$ the result of our self-consistent calculation. Some $J_{ij}$ in our calculation are ferromagnetic (but smaller than the corresponding antiferromagnet interactions at similar distances). These reduced-strength sites seem not to be numerical artefacts, as they appear for different system sizes and numerical regularisations. They also match the lattice symmetries despite the system boundaries not doing so. Broadly, though, the gapped graphene analytical function remains a decent rule of thumb for the $J_{ij}$ in the $\mathbb{Z}_2$ QSL. We nevertheless use the results of our numerical calculations in the subsequent sections.

\subsubsection{Magnons}
\label{subsec:magnons}

Early theoretical studies on the KHAF calculated a spin-spin susceptibility using the Kondo-Yamaji Green's function method~\cite{Bernhard02,LeggBraunecker19}. These calculations are interesting for two reasons. First, the susceptibility can be attributed to magnons despite the lack of long-range magnetic order. Second, the calculation remains agnostic to the particular form of QSL ansatz, focussing instead on the microscopic spin interactions. 

\begin{figure}
    \centering
\includegraphics[width=\linewidth]{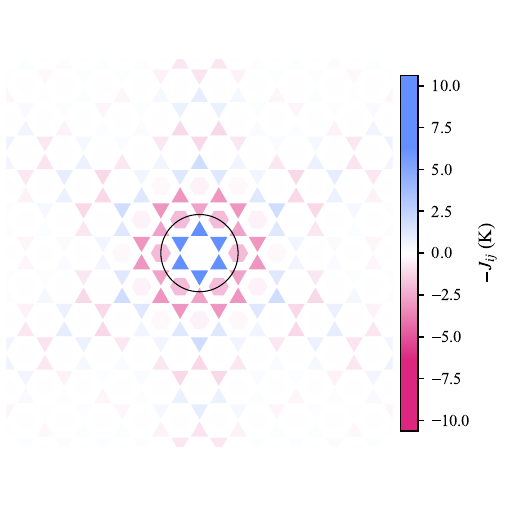}
    \caption{Calculated $J_{ij}$ for magnon-mediated interactions (Sec.~\ref{subsec:magnons}). The strength decays as $1/r^2$ in most directions, but as $1/r$ along high symmetry directions such as the $y-$axis. As a result, sums over these interactions are divergent.}
    \label{fig:J_ij_magnon}
\end{figure}

Reproducing this calculation, we converted the result to real-space $J_{ij}$ terms using a Fourier transform. The result is shown in Fig.~\ref{fig:J_ij_magnon}. We find that the interactions decay as $1/r^2$ in most directions, but decay as $1/r$ along certain high symmetry directions. These slow decays lead to divergent sums over spin interactions: the spins along a circle of radius $R$ around a given point contribute more for larger $R$. As well as being unphysical (the energy of such a set of interactions would diverge), this leads to an incorrect prediction for neutron scattering. This rules out a magnon-mediated witness interaction of this form as a possible explanation for the observations in Herbertsmithite.

%
\section{Monte Carlo Methods}
\label{sec:Methods}
%

\begin{figure}
    \centering
    \includegraphics[width=0.6\linewidth]{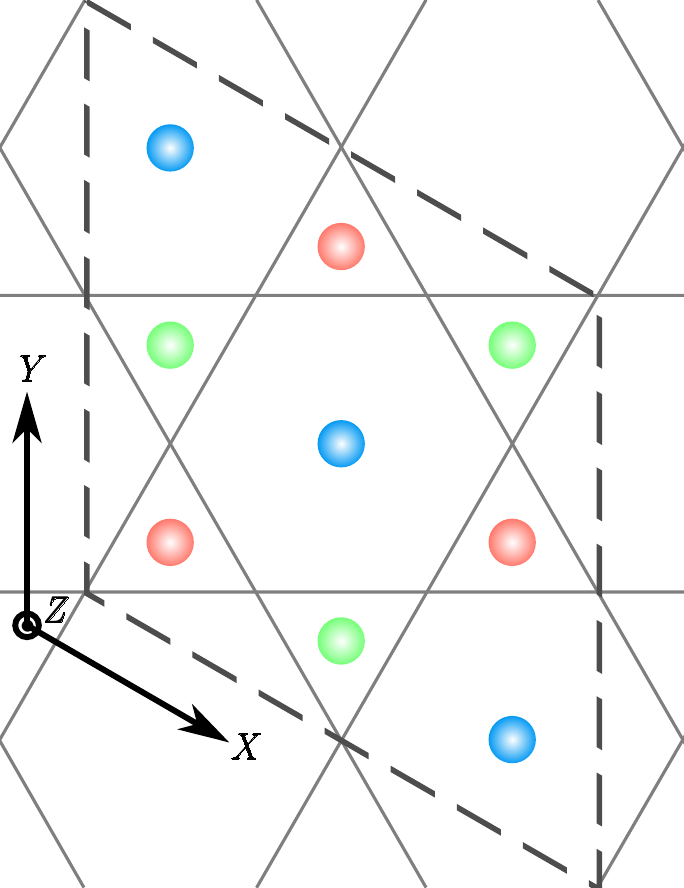}
    \caption{Diagram showing a 3 x 3 x 1 cell (dashed line) in the simulated lattice. Each $Z$ layer contains three witness planes to capture the ABC stacking. The circles indicate potential witness sites and the color indicates the witness plane of the site (\emph{i.e.} red is in the A-plane, green is in the B-plane, blue is in the C-plane.) The solid lines illustrate the kagome layer between A-plane and B-plane (not part of the simulated lattice).}
    \label{fig:simulation}
\end{figure}

Having identified the couplings $J_{ij}$ between witnesses in Sec.~\ref{sec:Model}, we proceed to simulate their dynamical and equilibrium behaviours. In order to simulate sufficiently large system sizes and run times to capture spin glass behaviour, from here on we approximate the witnesses to be classical Ising spins (nominally aligned along $c$). We then create a classical Monte Carlo simulation of the witness spins using the Metropolis-Hasting algorithm~\cite{NewmanBarkema}.

The physical space represented by the simulation is a rhombic prism with the $xy$-planes forming the rhombic cross-section, as illustrated by Fig.~\ref{fig:simulation}. We indicate the size of the simulated lattice using the following notation: $X\times Y\times Z$ with $Y=X$. Note that this space has a different orientation to the usual crystallographic Herbertsmithite unit cell as defined in Ref.~\onlinecite{BraithwaiteEA04}. In this notation, $X$ indicates the side length of the rhombic cross-section equal to the number of witnesses and each integer $Z$ corresponds to a single period of the ABC stacking (\emph{i.e.} three layers of the witness plane). The side length and height of the rhombic prism are, respectively, $X\frac{a}{\sqrt{3}}$ and $Zc$, where $a = 6.834\,$\AA~and $c = 14.075\,$\AA~are Herbertsmithite's unit cell parameters~\cite{BraithwaiteEA04}. We further take the simulated lattice to have periodic boundary condition in all directions. As each $Z$ layer contains three witness planes, we require that $X$ is a multiple of three. Furthermore, although not a strict requirement, we only simulate lattices with even $Z$. As the models we are simulating include predominantly AFM interactions between nearest witness planes, we expect that a simulated lattice with odd $Z$ and periodic boundary condition could exhibit artificial frustration along the $z$-axis. For the results reported in this paper, we use $X = 45$ and $Z = 4$ for our simulated lattice. An exception to this is when the spins exhibit long-range ordering, where we use $X = 48$ in order to avoid suppressing the ordering.

To start a simulation run, we randomly assign $N_w = xX^2Z$ of the sites to be occupied by Cu${}^{2+}$ witnesses. The location of the witnesses is assigned using a seeded random number generator to allow us to isolate behaviours that are the result of a specific witness configuration versus a general behaviour of the model. All runs where $x<1$ are performed on 128 different seeds that are kept identical between runs. 

To thermalise and simulate the system at thermal equilibrium, we update it through a Monte Carlo (MC) sweep. A single MC sweep involves $N_w$ spin flip attempts, where each spin flip attempt involves picking a random witness and reversing the spin ($1\rightarrow-1$ and vice versa). To ensure that detailed balance holds, each spin flip attempt has $\text{max}(e^{-\Delta E/k_BT},1)$ chance of being accepted, where $\Delta E$ is the change in energy due to the spin flip.

Once the configuration of witness sites has been specified, we initialise the system with randomly oriented spins. We then run our simulation starting at an initial temperature $T_i\gg T^*$ and ending at final temperature $T_f$ using steps of $T_{step}$. At each temperature, the system is first equilibrated through $N_{eq}$ MC sweeps. Following equilibration, we then perform $N_T$ MC sweeps to obtain our samples for calculating physical observables. For our results, we used $N_{eq} =$ 2,000 MC sweeps and $N_T =$ 100,000 MC sweeps. 

Here, the thermal averaging $\langle \cdots \rangle$ indicates an average of the corresponding quantity over all MC sweeps performed at thermal equilibrium. Implicitly, all reported observables are additionally averaged over disorder---being the 128 different seeded witness configurations---at the final step.
    
We begin with the DC susceptibility $\chi(T)$ (not to be confused either with the static spin susceptibility $\chi^0$ or with the spinon hopping parameters $\chi_{lm}$), which we calculate from the fluctuations of the magnetization. We define the magnetization here as magnetization per spin,
\begin{align}
    m = \frac{1}{N_w}\sum_{i=1}^{N_w} s_i,
\end{align}
where $s_i=\pm1$ for Ising spins. The DC susceptibility (per spin) is obtained as the variance of the magnetization,
\begin{align}
    \chi = \frac{N_w}{T}\left(\langle m^2\rangle - \langle m\rangle^2\right).
\end{align}
We use the location of the cusp in $\chi$ to identify the glass transition temperature $T^*$.

We use the Edwards-Anderson order parameter $q_{\textrm{EA}}$ to characterise spin freezing. This quantity measures whether the spins in the configuration, on average, remain in the same orientation over time. It is defined as 
\begin{align}
    q_{\textrm{EA}} = \frac{1}{N_w}\sum_{i=1}^{N_w}\langle s_i\rangle^2.
\end{align}

A shortcoming of the Edwards-Anderson order parameter is that it also goes to unity for any phase where the spins become static, such as an AFM phase, and not just spin glass. In the case of spin glass, we expect that conventional real-space order parameters, such as the absolute magnetization $\langle|m|\rangle$ and the AFM order parameter $\langle q_{\textrm{AFM}}\rangle$, would remain nearly zero as per the paramagnetic phase. We defined $q_{\textrm{AFM}}$ as
\begin{align}
    q_{\textrm{AFM}} = \frac{1}{N_w}\left|\sum_{i=1}^{N_w} (-1)^\alpha s_i\right|,
\end{align}
where $\alpha = \pm 1$ is the sublattice index; \emph{i.e.} its sign alternates between $xy$-planes in real space.

We calculate the energy-integrated neutron scattering intensity $S(\mathbf{q})$ using
\begin{align}
S(\mathbf{q})=\left|F(\mathbf{q})\right|^2\left\langle\left|\sum_i s_i\exp\left(i\mathbf{q}\cdot\mathbf{r}_i\right)\right|^2\right\rangle
\end{align}
and the noise spectral density as
\begin{align}
S(f)=\left| \sum^{N_T}_{t}\sum^{N_w}_i s_i(t)\exp\left(2\pi ift\right)\right|^2.
\end{align}
Here $N_T$ is the number of sampling MC sweep in the simulation and $F(\mathbf{q})$ is the Cu$^{2+}$ atomic form factor~\cite{HanEA16}, which we use in our plots. As we do not have a physical timescale with which to convert Monte Carlo time steps into physical time units, the frequency $f$ is in arbitrary units. 

%
\section{Results}
\label{sec:Results}
%

%
\subsection{Neutron Structure Factor}
\label{subsec:neutrons}

\begin{figure*}
    \centering
    \includegraphics[width=0.99\linewidth]{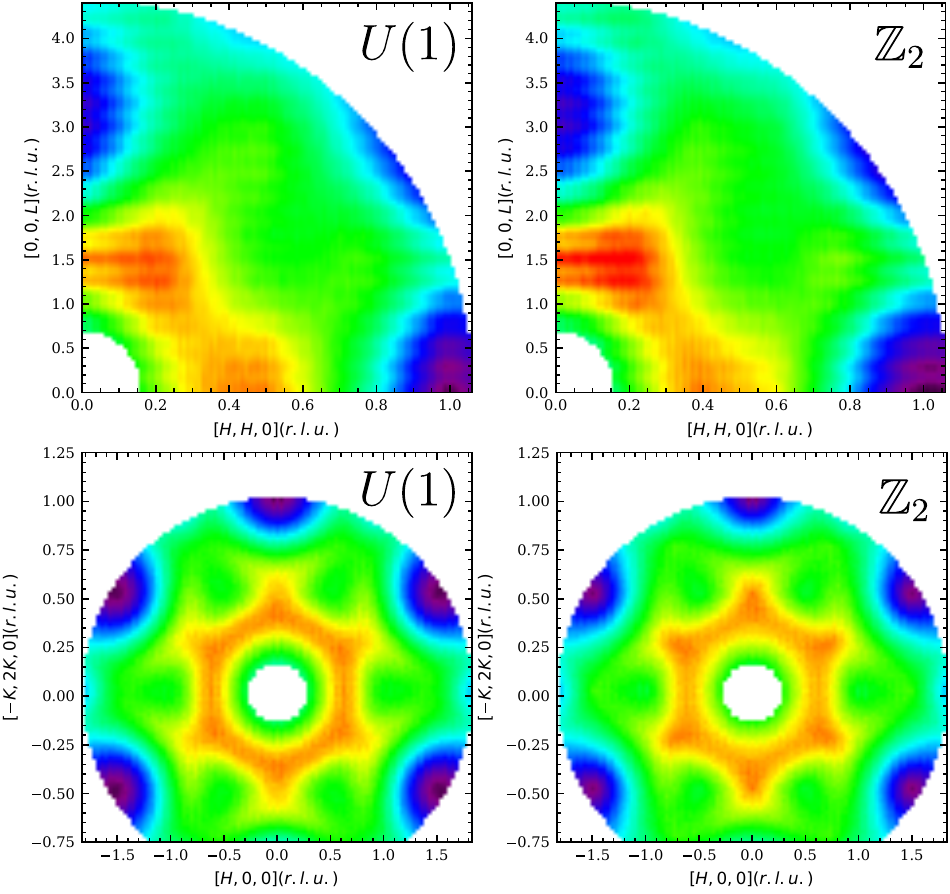}
    \caption{Our models' predictions of the energy integrated neutron scattering structure factors at $2\,$K. The two candidate QSLs are essentially indistinguishable. Plots share a common colour bar (intensity in arbitrary units). The broad features at $00\frac{3}{2}$ and $\frac{1}{2}\frac{1}{2}0$ were previously explained with a model of nearest neighbour AFM correlations between witnesses~\cite{HanEA16}.}
    \label{fig:S_Q_2K}
\end{figure*}

In Fig.~\ref{fig:S_Q_2K}, we plot our predicted static/energy-integrated neutron scattering intensity $S(\mathbf{q})$ at $T=2\,$K. The results can be compared directly with the experimental and theoretical results in Fig.~1 of Ref.~\onlinecite{HanEA16}. We use the same form factor as indicated in that work. As noted in that paper, the key features are a diffuse ring of scattering in the HK0 plane, and broad peaks (not diffraction limited) in the HHL plane at $00\frac{3}{2}$ and $\frac{1}{2}\frac{1}{2}0$. Both the $\mathbb{Z}_2$ and $U(1)$ QSLs reproduce these features well. Ref.~\onlinecite{HanEA16} shows that the HHL peaks can be reproduced by a simple model of AFM correlations, without long-range order, between nearest neighbour witnesses (which are displaced along $c$). 

The experimental HHL scattering pattern, in particular, places a strong constraint on possible interactions between witnesses. It suggests that (i) the nearest neighbour witness-witness interactions should be antiferromagnetic; (ii) further neighbour interactions must exist to frustrate the formation of long-range order; (iii) the strengths of further neighbour interactions must decay sufficiently quickly with distance that other features do not appear. The necessity of (ii) is clear from the fact that the nearest neighbour structure alone has the connectivity of the simple cubic lattice; it is therefore bipartite and unfrustrated with AFM interactions. 

The necessity of (iii) is demonstrated clearly by the non-QSL model of Sec.~\ref{subsec:magnons}, in which witness interactions are mediated by magnons. In this case the couplings $J_{ij}$ are found to obey criteria (i) and (ii), but the AFM couplings decay with distance $r$ as $1/r^2$, or even $1/r$ along high symmetry directions. As a result, sums over spins are divergent: the sum of contribution of spins at radius $R$ grows with $R$. One effect of this is that the neutron scattering pattern cannot match experimental observations, and such a model can be neglected. 

The neutron scattering data therefore constrain possible models of witness interactions as follows: interactions must be antiferromagnetic at nearest neighbour; frustrated by further neighbour interactions; and must decay at least as fast as $1/r^2$ (which is in any case a physical requirement for quasi-2D systems). Fig.~\ref{fig:S_Q_2K} shows that both QSL candidates in Herbertsmithite meet these criteria. Unfortunately, they lead to essentially indistinguishable scattering patterns at $2\,$K, meaning the existing neutron data cannot discern between the two possibilities.

\begin{figure*}
    \centering
\includegraphics[width=0.99\linewidth]{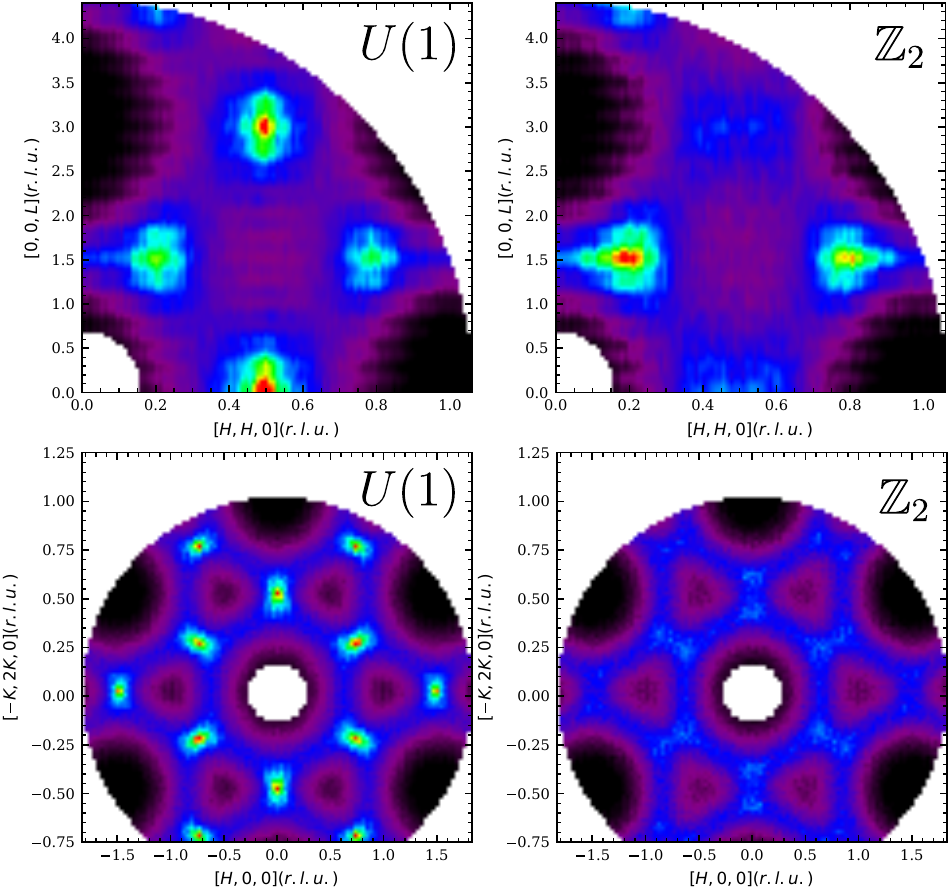}
    \caption{Energy-integrated neutron scattering structure factor predictions at 100 mK. Plots share a common colour bar (intensity in arbitrary units). The new feature at $\frac{1}{2}\frac{1}{2}3$ in the $U(1)$ QSL, not present in the $\mathbb{Z}_2$ QSL, would allow an unambiguous differentiation between the two cases.}
    \label{fig:S_Q_100mK}
\end{figure*}

However, turning our attention to possible near term experiments, in Fig.~\ref{fig:S_Q_100mK} we show the predicted neutron scattering patterns at $100\,$mK, within the spin glass regime for both models. In both QSLs, the $00\frac{3}{2}$ feature has sharpened and in so doing has moved towards $\frac{1}{5}\frac{1}{5}\frac{3}{2}$. We further observe a qualitative difference between the two candidate QSLs. The ratio $S(\frac{1}{2}\frac{1}{2}0)/S(\frac{1}{5}\frac{1}{5}\frac{3}{2})$, which was around 1 at $2\,$K, has decreased to around zero for the $\mathbb{Z}_2$ QSL, but for the $U(1)$ QSL it has actually increased to greater than one. A new feature has also appeared in the $U(1)$ pattern at $\frac{1}{2}\frac{1}{2}3$ which again outweighs the $\frac{1}{5}\frac{1}{5}\frac{3}{2}$ peak; this feature is absent in the $\mathbb{Z}_2$ pattern. These clear differences between the neutron scattering predictions for the two QSLs at low temperatures offer a route to definitively identifying which is present in Herbertsmithite. 

It can be helpful to imagine starting from the long-range order observed in the $x=1$ case (see Sec.~\ref{subsec:x}) and randomly removing sites. The neutron scattering at the physical concentration can then be thought of as residual ordering in local witness clusters. For instance, the diffuse peak for $U(1)$ at $\frac{1}{2}\frac{1}{2}0$ and $\frac{1}{2}\frac{1}{2}3$, corresponds to nearest neighbour ferromagnetic correlation. Although this may appear counterintuitive as all couplings for the $U(1)$ model are antiferromagnetic, the same correlation can also be observed in the long-range order phase for $U(1)$ in Sec.~\ref{subsec:x}. This suggests that the spin glass in our model can be thought of as glassy short-range ordered regions where each region exhibits similar correlation to the long-range order phase.

\subsection{DC Magnetic Susceptibility}
\label{subsec:chi_T}

\begin{figure}[t]
    \centering
    \includegraphics[width=0.99\linewidth]{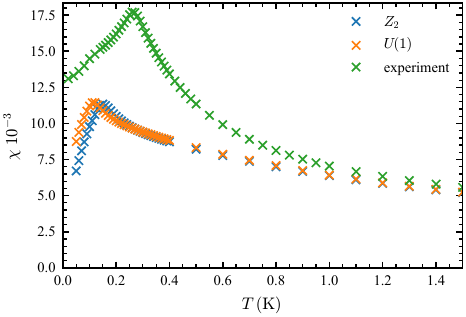}
        \caption{DC magnetic susceptibility $\chi(T)$. Green: experimental data reproduced from Ref.~\onlinecite{TakahashiEA26}. Blue, orange: numerical results for our model with $\mathbb{Z}_2$ and $U(1)$ QSL, respectively. We fix our free parameter $\gamma$ to match experimental $T_{\textrm{CW}}$.}
    \label{fig:chi_T}
\end{figure}

\begin{figure}[b]
    \centering
    \includegraphics[width=0.99\linewidth]{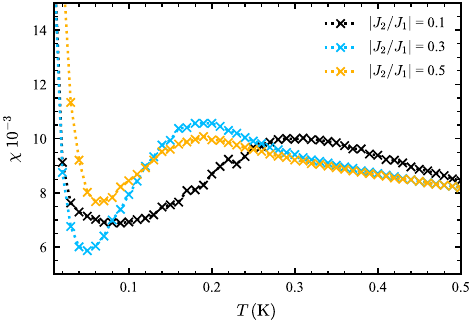}
    \caption{DC magnetic susceptibility $\chi(T)$ for a witness model with only nearest neighbour coupling $J_1$ and next nearest neighbour coupling $J_2$. All couplings are antiferromagnetic. The divergent susceptibility at low $T$ demonstrates that couplings to further neighbours are required for a spin glass.}
    \label{fig:X_NNN}
\end{figure}

In Fig.~\ref{fig:chi_T} we reproduce the measured DC magnetic susceptibility data $\chi(T)$ from Ref.~\onlinecite{TakahashiEA26}, showing the cusp at $T^*=260\,$mK. We also show our predictions based on both $\mathbb{Z}_2$ and $U(1)$ QSLs. In both cases we have constrained the one free parameter in our model, the witness-kagome coupling $\gamma$, to give a match to the observed Curie-Weiss temperature of $T_{\textrm{CW}}=1.1\,$K. Experimentally, $T_{\textrm{CW}}$ is found by a linear fit to $\chi(T)^{-1}$ at high temperature extrapolated to find the negative of the $x-$intercept. As a result, constraining to the Curie-Weiss temperature causes our models to agree with the experimentally observed $\chi(T)$ at high temperatures. We note that constraining $\gamma$ does not affect the ratio $T_{\textrm{CW}}/T^*$. 

We find that the $\mathbb{Z}_2$ QSL gives $T^*=0.15\,$mK while the $U(1)$ QSL gives $T^*=0.11\,$K. In both cases $T^*$ is heavily suppressed compared to $T_{\textrm{CW}}=1.1\,$K, as in the experiment. This is a measure of the frustrating effect of further-neighbour interactions. The higher value of $T^*$ in the $\mathbb{Z}_2$ QSL compared to $U(1)$ can similarly be understood as relatively less frustration, since $J_{ij}$ decay exponentially for $\mathbb{Z}_2$ and only algebraically, as $1/r^3$, for $U(1)$. The $\mathbb{Z}_2$ prediction is closer to the experimental value, but the approximations in our model (notably the classical Ising nature of the witnesses, and the use of mean field theory in the QSL calculations) likely negate the possibility of drawing any inferences from these small differences. 

We also show in Fig.~\ref{fig:X_NNN} the result of a model that includes only nearest and next nearest neighbour witness interactions. The divergence at low $T$ is representative of the behaviour of any such local interaction models. It can be understood as follows. Even though the majority of witnesses form a single large connected cluster with second-nearest neighbour interactions and $x=33$\% site occupation, a significant minority of witnesses are isolated or are member of small clusters whose interaction is dominated by frustration from the next nearest neighbour coupling. These behave paramagnetically, giving a divergent susceptibility at low $T$ that ultimately dominates the signal. Any models based only on near neighbour witness exchange will fail to form a connected cluster across the bulk, adding another constraint on possible models. For this reason, even if appropriate witness exchange pathways could be identified in Herbertsmithite, an explanation of the experimental data would be unlikely to succeed even with significant fine tuning. 

It is notable that, even though the interactions induced by the $\mathbb{Z}_2$ QSL decay exponentially with a decay range of less than one lattice site, the exponential tail of this interaction nevertheless succeeds in connecting all witnesses into a single connected component, thus avoiding a divergent $\chi(T)$ at low temperatures.

\begin{figure}[t]
    \centering
    \includegraphics[width=0.99\linewidth]{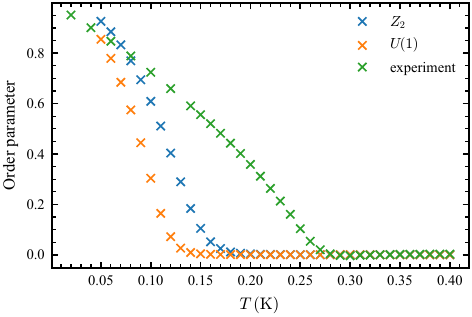}
    \caption{Edwards-Anderson order parameter $q_{\textrm{EA}}(T)$. For both QSLs, $q_{\textrm{EA}}$ changes smoothly near $T^*$. This matches the behaviour of the experimental values taken from Ref.~\onlinecite{TakahashiEA26} (green).}
    \label{fig:q}
\end{figure}

\begin{figure*}[ht!]
    \centering
    \subfigimg[width=0.49\linewidth]{}{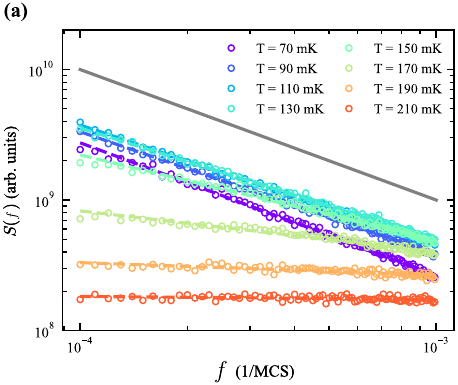}
    \hfill
    \subfigimg[width=0.49\linewidth]{}{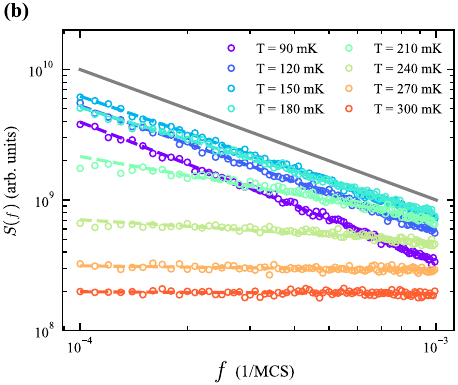}
    \caption{Power spectral density for (a) $U(1)$ and (b) $\mathbb{Z}_2$. Both show white noise above $T^*$ and $1/f$ `pink' noise below $T^*$, indicative of a spin glass. The grey solid lines correspond to $1/f$ noise.}
    \label{fig:PSD}
\end{figure*}

\subsection{Spin Glass Order Parameter}
\label{subsec:q}

In Fig.~\ref{fig:q} we show the Edwards-Anderson spin glass order parameter $q_{\textrm{EA}}$. We show the experimental values of $q_{\textrm{EA}}$ from Ref.~\onlinecite{TakahashiEA26}, which were obtained in that paper using
\begin{align}
\chi(T)\propto\frac{1-q_{\textrm{EA}}(T)}{T-T_{\textrm{CW}}\left(1-q_{\textrm{EA}}(T)\right)}.
\end{align}
The experimental data show $q_{\textrm{EA}}$ turning on smoothly below $T^*$, the temperature of the cusp in $\chi(T)$. Our QSL predictions mirror this behaviour, with the respective $T^*$ seen in the models. Additionally, we observe $\langle q_{\textrm{AFM}}\rangle=0$ in our models, ruling out the possibility of long-range AFM order. The neutron scattering is not featureless, however, suggesting that the spin glass possesses localised AFM correlated regions. We return to this in Section~\ref{subsec:x}. 

\subsection{Noise Spectral Density}
\label{subsec:PSD}

In Fig.~\ref{fig:PSD} we show the magnetization noise spectral density $S(f,T)$ as a function of frequency $f$. Again matching the experimental data of Ref.~\onlinecite{TakahashiEA26}, we find white noise ($S$ independent of frequency) for $T>T^*$ which turns over to $S\propto 1/f$ `pink' noise below $T^*$. This is a well-documented behaviour of spin glasses~\cite{OcioBouchiatMonod85,Weissman93,AtalayaEA14}.

\subsection{Varying Witness Concentration}
\label{subsec:x}

Our model suggests that the magnetic impurities in Herbertsmithite can act as a direct probe of the properties of the QSL~\cite{TakahashiEA26}. While earlier studies were motivated to reduce the concentration of impurities, seeing them as an obstacle to observing the QSL, this new perspective begs the question as to how differing impurity concentration $x$ can affect the properties of Herbertsmithite. 

Reported impurity concentrations for Herbertsmithite vary substantially between analyses and samples. A recent study reported 11\% Zn-in-kagome, but with nearly stoichiometric Zn:Cu\cite{KremerEA25}. Earlier work reported a synthesis that resulted in stoichiometric Herbertsmithite with 7\% antisite disorder as well as another sample that is Cu-deficient, with 11\% of Cu site occupied by Zn ions~\cite{deVriesEA12}. Another study reported 15\% of Zn site occupied by Cu with no significant Zn occupation in Cu sites~\cite{FreedmanEA10}. In our model, this corresponds to a witness occupation of $x=0.15$. However, Ref.~\onlinecite{TakahashiEA26} revisited earlier X-ray diffraction data taken on the same crystals~\cite{KremerEA25}, and, based on the Curie-Weiss fitting, calculated the witness occupation to be $x=0.33$. Several experimental studies have reported that Zn substitution into Cu planes is unlikely~\cite{FreedmanEA10,HanChuLee12}, and on that basis it has been claimed that there is a minimum possible value of $x$ (then reported as $x=0.15$). In the opposite limit, it is known that a structural phase transition occurs at around $x=0.66$ where the crystal loses its undistorted kagome lattice structure~\cite{Norman16}. Given these constraints, we expect that our theoretical model is only applicable to Herbertsmithite with impurity concentration between $0.15<x<0.66$. However, owing to finite system sizes in our simulation, the precise values of $x$ we report are likely inaccurate, although we would expect that the broad trends would match what would be seen. To better understand the behaviour of our model, we thus conducted numerical calculations for $0\leq x\leq 1$.

\begin{figure*}
    \centering
        \subfigimg[width=0.95\linewidth]{}{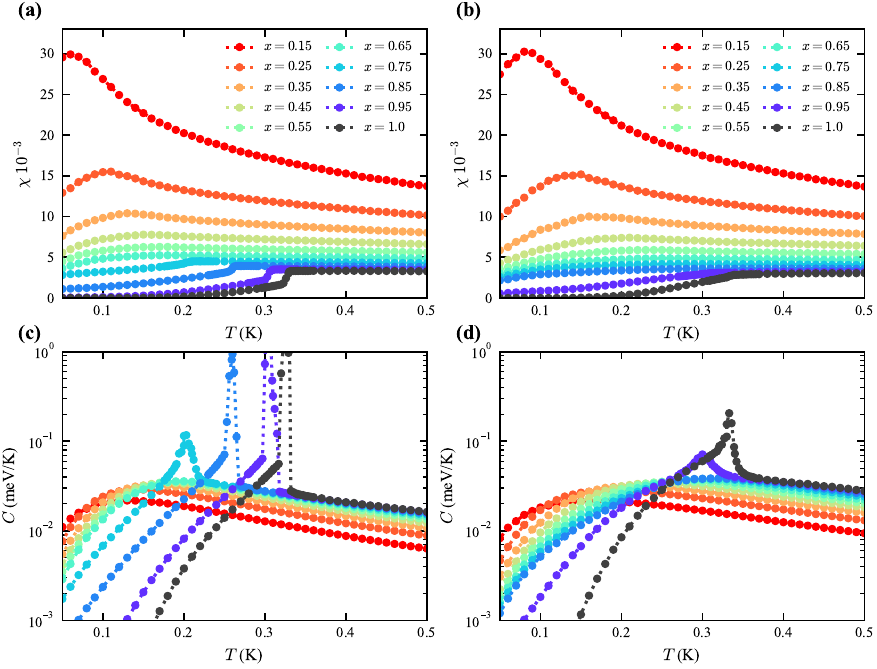}
    \caption{DC witness susceptibility $\chi(T)$ for various witness concentration $x$ for (a) $U(1)$ and (b) $\mathbb{Z}_2$. In both model, the susceptibility is largely symmetric around the peak at low $x$. At $x\geq0.75$ for $U(1)$ and $x\geq0.95$ for $\mathbb{Z}_2$, $\chi(T)$ shows an unsymmetrical drop off for $T < T^*$. A corresponding change occurs in witness specific heat $C(T)$ for (c) $U(1)$ and (d) $\mathbb{Z}_2$. At the same threshold, a change in behaviour is also observed where the smooth peak becomes increasingly asymptotic. These phenomena are indicative of a change from spin glass transition to LRO transition. The result for the LRO regime was performed using simulated lattice size of 48x48x4 instead of 45x45x4.}
    \label{fig:chi_T_x}
\end{figure*}

In Fig.~\ref{fig:chi_T_x} we show the evolution of the DC magnetic susceptibility $\chi(T)$ and specific heat $C(T)$ with varying $x$ for both QSLs. As $x$ increases, the cusp in $\chi(T)$ moves to higher temperatures: higher witness concentration means higher witness density and therefore overall stronger interactions. We further observe that, at high $x$, the specific heat develops a sharp step. This is indicative of a phase transition to long-range ordering rather than spin glass. This fits the intuition that the lack of disorder at $x=1$ means that we expect some form of ordered state at low temperature rather than spin glass. 

\begin{figure}
    \centering
        \subfigimg[width=0.49\linewidth]{}{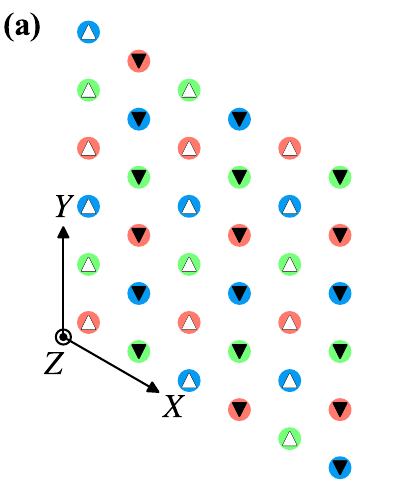}
        \subfigimg[width=0.49\linewidth]{}{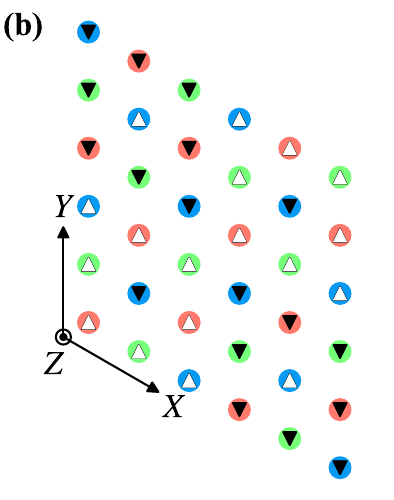}
    \caption{Schematic picture of the form of long-range order in real space for (a) $U(1)$ and (b) $\mathbb{Z}_2$ across the $xy$-plane. Not shown is the ordering across lattice unit cell along the $z$-axis, where (a) is ferromagnetic and (b) is antiferromagnetc. Red sites are in the A-plane, green sites in the B-plane, and blue sites are in the C-plane. White up-triangle indicates spin-up and black down-triangle indicates spin-down.}
    \label{fig:LRO}
\end{figure}

We carried out numerical simulations for $0\leq x \leq 1$ at intervals of $x = 0.05$. These results indicate that a first-order phase transition occurs at sufficiently large $x\approx0.95$ for $\mathbb{Z}_2$ and $x\approx0.75$ for $U(1)$. At these concentrations, enough witnesses exist to make long-range ordering energetically favourable. The LRO pattern differs between the two QSLs; their ordering is illustrated in Fig.~\ref{fig:LRO}. In both cases, the LRO phase shows a striped pattern in the $xy$-plane with 3-fold degeneracy up to translational and spin inversion symmetry. This result is consistent with previous studies of AFM models on the triangular lattice (recalling that witnesses occupy a triangular lattice between kagome layers) where long-range interactions result in stripe order\cite{Korshunov05}. In particular, Ref.~\onlinecite{KoziolEA23} demonstrates that AFM coupling with power law decay generally exhibits stripe ordering on the triangular lattice. This is consistent with the ordering pattern of $U(1)$ QSL model on the same witness plane.

\begin{figure}[b!]
    \centering
    \begin{subfigure}{0.99\linewidth}
        \centering
        \includegraphics[width=\linewidth]{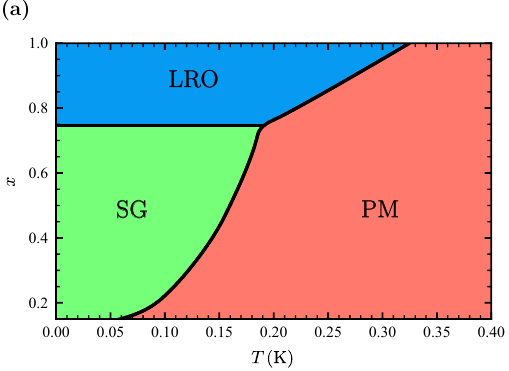}
    \end{subfigure}
    \begin{subfigure}{0.99\linewidth}
        \centering
        \includegraphics[width=\linewidth]{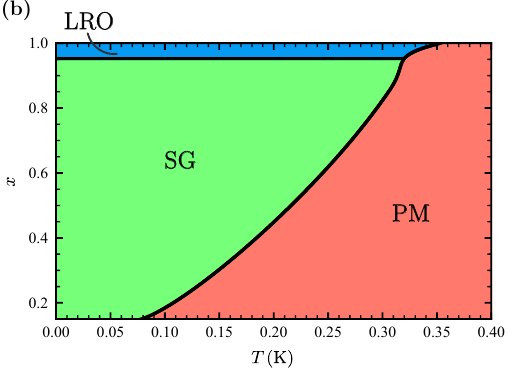}
    \end{subfigure}

    \caption{Phase diagrams for (a) U1 and (b) Z2. PM: paramagnet; SG: spin glass; LRO: Long-range order.}
    \label{fig:phase_diagrams}
\end{figure}

Using these data, in Fig.~\ref{fig:phase_diagrams} we construct phase diagrams for both the $\mathbb{Z}_2$ and $U(1)$ QSLs as a function of $x$ and $T$. At low witness concentration $x<0.15$, we find the paramagnetic phase freezing to spin glass as indicated by $q_{\textrm{EA}}$ approaching unity as $T\rightarrow0$. However, we are not able to report the precise $T^*$ due to a lack of cusp in $\chi(T)$. At $x > 0.15$, we identify a clear spin glass transition temperature $T^*$. The quantity reported in this diagram shows the average with respect to site disorder with a standard deviation of $20\,$mK. 

Though the phase diagrams for both QSLs are qualitatively similar, we can see the effect of the interaction range on the phase diagram. We can see that the LRO phase is more vulnerable to missing-site disorder for the  $\mathbb{Z}_2$ QSL's exponentially decaying $J_{ij}$, compared to the $U(1)$ QSL's algebraic $1/r^3$ decay. Short-range interactions mean that local ordering of sites is much more vulnerable to being disrupted by missing sites. Our models predict that the threshold between spin glass and LRO phase sits outside of the physically meaningful range for Herbertsmithite. However, if this transition were observed in higher impurity samples, it would be indicative of long-range coupling between witnesses.

%
\section{Conclusions}
\label{sec:conclusions}
%

We have detailed our model, presented in part in Ref.~\cite{TakahashiEA26}, of spin glass formation amongst spin-1/2 Cu${}^{2+}$ impurities (`witnesses') on Zn${}^{2+}$ sites in Herbertsmithite as originating from interactions mediated by the quantum spin liquid (QSL) in the kagome copper layers. Our model has only one free parameter: the strength of witness-kagome coupling $\gamma$ which we constrain to $\gamma=\pm60\,$K by requiring a Curie-Weiss temperature of $T_{\textrm{CW}}=1.1\,$K. Despite this, we find a good quantitative and qualitative match to the full range of existing experimental data, including: a cusp in the DC magnetic susceptibility $\chi(T)$ at a temperature $T^*$ well below $T_{\textrm{CW}}$; the smooth turning on of an Edwards-Anderson order parameter $q_{\textrm{EA}}$ at $T^*$; the full frequency and temperature dependence of the noise spectral density $S(f,T)$; and the elastic neutron scattering intensity $S(\mathbf{q})$ in both the HK0 and HHL planes at $2\,$K. We have additionally made new predictions for future experiments, including a phase diagram for the predicted behaviour as a function of witness concentration $x$. Importantly, we find that if elastic neutron scattering can be undertaken below $260\,$mK, the two candidate QSLs, $\mathbb{Z}_2$ and $U(1)$, can readily be distinguished owing to the appearance of a strong broad peak at $\frac{1}{2}\frac{1}{2}3$ in the $U(1)$ QSL, not present in the $\mathbb{Z}_2$ QSL. By considering the long-range order at $x=1$, which again differs between $U(1)$ and $\mathbb{Z}_2$, we found that the $x=0.33$ can be understood in terms of short-range ordered regions with glassy dynamics. This explains why previous neutron scattering measurements at $2\,$K saw some structure (albeit far from diffraction limited), where the simplest expectation for a spin glass would be an isotropic ring of scattering~\cite{PhysRevB.89.054433}. 

\subsection{Limitations of our approximations}
\label{subsec:limitations}

While both the $\mathbb{Z}_2$ and $U(1)$ QSLs reproduce the full range of existing experimental data, $\mathbb{Z}_2$ does slightly better in the sense that its predicted $T^*$ is closer to the experimental value (leading to a closer match to the $\chi(T)$, $S(f,T)$, and $q_{\textrm{EA}}$ predictions). We do not read too much into this success, since some of our approximations are rather drastic and likely outweigh this small difference in predictions. Two approximations in particular could be removed with further work.

First, we treat the QSLs using mean field theory (self-consistent mean field theory for the $\mathbb{Z}_2$ QSL; no self-consistency is required for $U(1)$). A more accurate estimate of the interactions between witnesses could be obtained using exact diagonalisation or DMRG. This would naturally be at the expense of considering a reduced system size compared to ours (we considered up to $32\times 32$ unit cells). We note that there remains significant disagreement over the nature of the ground state of the KHAF, with some studies reporting the $U(1)$ QSL to be of lower energy, and others reporting $\mathbb{Z}_2$. Recent advances in machine learning approaches have led to a report that in fact the ground state is not a QSL at all, but is instead a spinon spin density wave. In order to obtain the witness couplings $J_{ij}$, these numerical methods would need to introduce a degree of three dimensionality, enhancing the issue. 

Second, after integrating out the QSL, we then treat the witnesses as classical spins. This approximation is largely pragmatic as the system needs to be sufficiently large to get a sensible measure of spin glass behaviour. Being spin-1/2, the witnesses are maximally quantum, so the classical approximation is likely to present inaccuracies. Given the $J_{ij}$ we have calculated, the relative sparseness of witnesses could make exact diagonalisation or DMRG plausible for estimating the resulting quantum witness behaviour. This would be highly desirable, since in principle the witnesses should inherit the long-range entanglement from the QSL. In this sense, the spin glass the witnesses form is really a new state of quantum matter: a long-range entangled spin glass. This could perhaps be detectable via measures such as the quantum Fisher information extracted from inelastic neutron scattering. 

We further justify the approximation of the spins as classical Ising spins rather than Heisenberg spins. From a numerical perspective, the spin glass state was first explicitly demonstrated in an Ising spin system~\cite{BinderStauffer76,BinderSchroeder76}. Several early numerical studies with Heisenberg spin system had, initially, failed to produce a spin glass transition at finite temperature. Large scale computational study eventually produces phenomena consistent with spin glass state in isotropic Heisenberg spin models~\cite{BanavarCieplak82,McMillan85,OliveEA86}. However, spin glass behaviour remains substantially more difficult to establish in these cases owing to finite-size effects~\cite{FernandezEA09}. Study of models with random anisotropic interaction are capable of producing spin glass behaviour more decisively, but these spin glass states were found to be in the same universality class as the Ising spin glass~\cite{MatsubaraEA91,BaityJesiEA14}. This suggests that Heisenberg spin glass can be represented by an effective Ising model as in our case. 

The Dzyaloshinskii-Moriya (DM) interaction is likely to be significant for witnesses in  Herbertsmithite~\cite{ZorkoEA08,HeullyAlary2025IsHF}. This could favour either an easy axis (Ising-like spins) along $c$, or an easy plane (XY-like) perpendicular to it. A model properly incorporating this effect would be desirable. 

\subsection{Alternative Models}
\label{subsec:other_models}

Both the $\mathbb{Z}_2$ and $U(1)$ models give similar matches to existing experimental data. This begs the question as to whether other models, not involving QSLs, might also be compatible with current observations.

We can make certain clear statements on the possible forms of witness-witness interactions $J_{ij}$ consistent with current experiments. (i) Neutron scattering at $2\,$K shows that the nearest neighbour witness-witness interactions are antiferromagnetic and the dominant interaction~\cite{NilsenEA13,HanEA16}. Spin glass formation at $T^*=260\,$mK, evidenced by $\chi(T)$, $S(f,T)$, and aging, shows that (ii) frustrating further neighbour interactions must exist; otherwise the lattice is trivially antiferromagnetic. The degree of frustration can be quantified by $T_{\textrm{CW}}/T^*=4.2$. This adds a significant constraint on possible couplings according to the mean-field expression 
\begin{align}
T_{\textrm{CW}}=\frac{S(S+1)}{3}\sum_i n_i J_{i0}
\end{align}
where $S=1/2$ and $n_i$ is the number of $i^{\textrm{th}}$ nearest neighbours to witness site $j=0$. (iii) In order for these further neighbour couplings not to destroy the match to neutron scattering, the interactions must decay at least as fast as $1/r^2$ (if a 2D or semi-2D model such as ours is considered; otherwise $1/r^3$ for a 3D interaction model). (iv) In addition, if near-neighbour magnetic exchange models are proposed, they must maintain sufficiently long-range exchange that the witnesses not present in the largest connected component do not lead to a divergent $\chi(T)$ at low temperatures. Ref.~\onlinecite{TakahashiEA26} measured down to $20\,$mK and saw no evidence for any upturn in $\chi(T)$. (v) Any local exchange models must also not induce a significant AFM order parameter: even if this order parameter does not correspond to true long-range AFM order, its development at a specific temperature would lead to clear signatures in specific heat that are not seen~\cite{HeltonEA07, deVriesEA08, KimchiEA18, MurayamaEA22}.

Together, constraints (i)-(v) significantly restrict any possible proposals for alternatives to QSL mediated interactions, even before any appeal is made to Occam's razor to dismiss highly fine-tuned interactions. For example, we considered a model of magnon-mediated interactions based on the Kondo-Yamaji Green's function decoupling~\cite{YuFeng00,Bernhard02,LeggBraunecker19}; this fell foul of (iii), subsequently failing to reproduce the observed neutron scattering pattern (Sec.~\ref{subsec:magnons}). We also conducted an extensive numerical search of local exchange models. First and second neighbours alone fall foul of (iv), giving a divergent $\chi(T)$ at low temperatures. Still, density functional theory predictions for direct witness exchange pathways would be welcome. Various other models were considered and ruled out in Ref.~\onlinecite{TakahashiEA26}. 

Ultimately, while restrictions (i)-(v) leave space for a range of possible witness-witness interactions, it remains non-trivial that the two successful QSL-based models we present meet all criteria. We suggest this lends considerable weight to the hypothesis that Herbertsmithite hosts a QSL within its kagome copper planes. We hope that our new prediction, that neutron scattering below $260\,$mK can distinguish the two candidate QSLs, will motivate further experiments.

\section*{Acknowledgments} The authors thank J.~Murphy, J.~C.~S.~Davis, S.~A.~Kivelson, and P.~Manuel for helpful discussions and related collaborations, S.~M.~Hayden for useful comments on the manuscript, H.~Takahashi for both, and S.~J.~Blundell for use of his computational resources. F.~F.~acknowledges support from the Engineering and Physical
Sciences Research Council, Grant No.~EP/X012239/1. This work was carried out using the computational facilities of the \href{https://www.bristol.ac.uk/acrc/}{Advanced Computing Research Centre, University of Bristol}.

\bibliography{reference}

\end{document}